\documentclass[10pt, twocolumn]{IEEEtran}

\usepackage{multirow}
\usepackage{booktabs}
\usepackage{longtable}

\usepackage{cite}

\usepackage{graphicx}
\usepackage{amsmath}
\usepackage{amsfonts}
\usepackage{amssymb}
\usepackage{stfloats}
\usepackage{arydshln}
\usepackage{xcolor}
\usepackage{subfigure}
\graphicspath{{figure/}}

\newtheorem{remark}{Remark}

\usepackage{amsfonts, amsmath, mathrsfs, amsbsy, amssymb, dsfont, color}
\usepackage{graphicx}

\newcommand\prefixtext[1]{%
  \ifvmode\else\\\@empty\fi
  \noalign{%
    \penalty0%
    \vbox{\mathstrut}%
    \penalty10000%
    \vskip-\baselineskip
    \penalty10000%
    \vbox to 0pt{%
      \normalbaselines
      \ifdim\linewidth=\columnwidth
      \else
        \parshape\@ne
        \@totalleftmargin\linewidth
      \fi
      \vss
      \noindent#1\par}%
      \penalty10000%
      \vskip-\baselineskip}%
      \penalty10000}

\newtheorem{theorem}{Theorem}
\newtheorem{lemma}{Lemma}

\newcommand{\qed}{\nobreak \ifvmode \relax \else
      \ifdim\lastskip<1.5em \hskip-\lastskip
      \hskip1.5em plus0em minus0.5em \fi \nobreak
      \vrule height0.75em width0.5em depth0.25em\fi}

\DeclareMathAlphabet{\mathpzc}{OT1}{pzc}{m}{it}
\newcommand{\comment}[1]{}
\def\({\left(}
\def\){\left)}

\def\bth{\boldsymbol{\theta}}

\def\bth{\boldsymbol{\theta}}

\def\bSi{\boldsymbol{\Sigma}}

\def\tr{\text{tr}}
\def\({\left(}
\def\){\right)}
\def\[{\left[}
\def\]{\right]}
\def\BEq{\begin{eqnarray}}
\def\EEq{\end{eqnarray}}
\def\BE*{\begin{eqnarray*}}
\def\EE*{\end{eqnarray*}}
\def\BA{\begin{array}}
\def\EA{\end{array}}

\def\Nn{\nonumber}
\def\0{\mathbf{0}}
\def\1{\mathbf{1}}

\def\A{\mathbf{A}}

\def\e{\mathbf{e}}

\def\r{\mathbf{r}}

\def\t{\mathbf{t}}

\def\and{\prefixtext{and}}

\def\diag{{\rm diag}}

\def\Im{\mathrm{Im}}
\def\Re{\mathrm{Re}}

\title{One-Bit Spectrum Sensing for Cognitive Radio}
\author{Pei-Wen Wu, Lei Huang,~\IEEEmembership{Senior Member,~IEEE},~David Ram{\'\i}rez,~\IEEEmembership{Senior Member,~IEEE},~Yu-Hang Xiao,~\IEEEmembership{Member,~IEEE} and Hing Cheung So~\IEEEmembership{Fellow, IEEE}
\thanks{\textcolor{blue}{This work has been submitted to the IEEE for possible publication.
Copyright may be transferred without notice, after which this version may
no longer be accessible.}

P.-W. Wu, L. Huang and Y.-H Xiao are with the State Key Laboratory of Radio
Frequency Heterogeneous Integration, Shenzhen University, Shenzhen 518060,
China (e-mail: wupeiwen2021@email.szu.edu.cn; dr.lei.huang@ieee.org; yuhangxiao@szu.edu.cn).}
\thanks{David Ram{\'\i}rez is with the Department of Signal Theory and Communications, Universidad Carlos III de Madrid, Madrid 28903, Spain, and also with the Gregorio Mara{\~n}{\'o}n Health Research Institute, Madrid 28007, Spain (email: david.ramirez@uc3m.es).}
\thanks{H. C. So is with the Department of Electrical Engineering, City University of Hong Kong, Hong Kong, China (e-mail: hcso@ee.cityu.edu.hk).}
\thanks{The work of D. Ram{\'\i}rez was supported by the Ministerio de Ciencia e Innovaci{\'o}n, jointly with the European Commission (ERDF), under grant PID2021-123182OB-I00 (EPiCENTER), and by The Comunidad de Madrid under grant Y2018/TCS-4705 (PRACTICO-CM). The work of Lei Huang was supported by the National Science Fund for Distinguished Young Scholars under Grant 61925108, and by the Joint fund of the National Natural Science Foundation of China and Robot Fundamental Research Center of Shenzhen Government under Grant  U1913203. }
}

\begin{document}

\input{epsf}
\date{}

\maketitle
\maketitle
\begin{abstract}

Spectrum sensing in cognitive radio necessitates effective monitoring of wide bandwidths, which requires high-rate sampling. Traditional spectrum sensing methods employing high-precision analog-to-digital converters (ADCs) result in increased power consumption and expensive hardware costs. In this paper, we explore blind spectrum sensing utilizing one-bit ADCs. We derive a closed-form detector based on Rao's test and demonstrate its equivalence with the second-order eigenvalue-moment-ratio test. Furthermore, a near-exact distribution based on  the moment-based method, and an approximate distribution in the low signal-to-noise ratio (SNR) regime with the use of the central limit theorem, are obtained. Theoretical analysis is then performed and our results show that the performance loss of the proposed detector is approximately $2$ dB ($\pi/2$) compared to detectors employing $\infty$-bit ADCs when SNR is low. This loss can be compensated for by using approximately $2.47$ ($\pi^2/4$) times more samples. In addition, we unveil that the efficiency of incoherent accumulation in one-bit detection is the square root of that of coherent accumulation. Simulation results corroborate the correctness of our theoretical calculations.

\end{abstract}

\begin{IEEEkeywords}
One-bit ADC, performance degradation, Rao's test, spectrum sensing.
\end{IEEEkeywords}

\begin{sloppypar}
\section{Introduction}

Spectrum sensing is a crucial prerequisite for the dynamic allocation of spectrum resources in cognitive radio (CR) networks, as it is responsible for finding vacant channels (a.k.a. spectrum holes)~\cite{sunchunhua2007WCNC,Luwei2013TCOM,Majun2008TWC,Luwei2012ITC,Huang2015TSP}. In many application scenarios, the task is to monitor wideband channels, which indicates that high-speed sampling is involved. However, traditional spectrum sensing methods that assume perfect quantization typically require high-precision quantization to achieve optimal performance. Such high-speed and high-precision sampling results in large energy consumption, which may not be practically feasible.

To address this problem, an effective method is to decrease the quantization accuracy, particularly by using only one bit~\cite{choi2016TCOMM}. One-bit analog-to-digital converters (ADCs) only require a single comparator to complete the sampling and quantization process, offering advantages such as high sampling rate, low hardware complexity, and low power consumption compared to high-precision sampling~\cite{walden1999IJSAC,murmann2015MSSC,Ali2020tcomm}. For example, at a sampling rate of $3.2$ GSPS/s, an 8-bit ADC sampling~\cite{Reyes2019TCSI} requires $105$ $m$Watts while the one-bit ADC sampling~\cite{bhumireddy2013ICCAS} consumes only $20$ $\mu$Watts. Furthermore, the performance loss of one-bit radar detectors proposed in~\cite{xiao2022tvt} is only $2$ dB (${\pi}/{2}$) at low signal-to-noise ratios (SNRs), which can be compensated via increasing the number of samples by a factor ${\pi}/{2}$. These merits motivate the application of one-bit sampling techniques to spectrum sensing.

Many one-bit detection problems assume the availability of prior information, such as noise power, channel parameter characteristics, and/or signal characteristics~\cite{Ali2020tcomm,ALi2016WCL,chengziyang2019ISPL,xiao2022tvt}. However, this work focuses on one-bit spectrum sensing in the absence of prior information, also known as blind spectrum sensing. In this case, the probability mass function (PMF) of the one-bit observations is the product of orthant probabilities, which does not have a closed-form expression~\cite{Bar2002TAES,orthant1964,craig2008new}. Therefore, numerical techniques are needed when designing detectors using standard methods, e.g.,  generalized likelihood ratio test (GLRT)~\cite{JUNFANG2013ISPL}. In addition, numerical methods result in higher computational time and costs, contradicting the original purpose for simple spectrum sensing~\cite{Yucek2009SURV}. Therefore, a closed-form detector is desired.

A closed-form one-bit eigenvalue moment ratio (EMR) detector, inspired by the EMR detector~\cite{Huang2015ITVT}, was proposed in~\cite{Zhaoyuan2021WCL}.
It was demonstrated that the one-bit EMR is $3$ dB inferior to the $\infty-$bit EMR. However, performance degradation of one-bit sampling was proven~\cite{van1966PROC} to be only $2$ dB when the SNR is low. The result has been further corroborated in one-bit detection by~\cite{xiao2022tvt,JUNFANG2013ISPL}, and other one-bit signal processing problems by~\cite{Host-Maden2000TSP,mezghani2007ISIT,viterbi2013principles,stein2018TSP,Papadopoulos2001TIT,Mezghani2020TIT}.
The increased performance loss is due to the fact that the one-bit EMR is obtained by initially stacking the real and imaginary parts of the one-bit complex observations, followed by computing the EMR of the corresponding real-valued covariance matrix,  neglecting the circularity property of the unquantized signals.

In this paper, we formulate a detector for one-bit observations following the rule of Rao's test and taking into account the circularity property to enhance performance. The result turns out to be the second-order EMR of the one-bit complex-valued sample covariance matrix, rather than the expanded real-valued covariance matrix as presented in~\cite{Zhaoyuan2021WCL}.

%Indeed, our focus is on the low SNRs scenario, as spectrum sensing is required to operate effectively and reliably under such conditions~\cite{chen2011JSAC}.

To verify that the lower bound of $2$ dB loss is met, we analyze the performance degradation by comparing the proposed one-bit Rao's test with its $\infty$-bit counterpart. To enable such a comparison, we derive approximate distributions of the proposed detector under pure noise and low SNRs, which yield results that can be compared to the asymptotic distribution derived in~\cite{xiao2017approximate}. In particular, the null distribution follows the same $\chi^2$ distribution, while the non-null distributions in the low SNRs are all non-central $\chi^2$ distributions, albeit with different non-centrality parameters. By examining the non-centrality parameters, we can effectively quantify the performance degradation. In higher SNR scenarios, where the approximation breaks, a near-exact Beta-approximation using the  moment-based method~\cite{Luwei2012ITC,Luwei2013TCOM,Huang2015TSP,Huang2015twc} is also provided.

The contributions of this paper are as follows:

\begin{enumerate}
\item We present a novel detector based on Rao's test for blind spectrum sensing utilizing one-bit ADCs. The proposed detector is formulated in a closed-form manner, eliminating the need for numerical optimization. Moreover, we demonstrate that the proposed detector is equivalent to the second-order EMR detector (or John's detector~\cite{john1971,ramirez2013TIT}) using complex-valued one-bit observations. Our detector outperforms the one-bit EMR detector~\cite{Zhaoyuan2021WCL}, which adopts the expanded real-valued covariance matrix.

\item We derive near-exact null and non-null distributions of the proposed detector, enabling the calculation of false alarm and detection probabilities. Additionally, approximate null and non-null distributions under low SNRs are obtained, simplifying performance comparison.

\item We prove that the performance loss of the proposed detector in low SNR environments is approximately $2$ dB ($\pi/2$) compared to the detector using $\infty$-bit ADCs, which is smaller than the $3$ dB performance loss reported in~\cite{Zhaoyuan2021WCL}. Moreover, this loss can be compensated for by increasing the number of samples of our detector by a factor of approximately $2.47$ ($\pi^2/4$).

\item Upon comparison with the findings in~\cite{xiao2022tvt}, we arrive at an intriguing conclusion: The efficiency of coherent accumulation in one-bit detection is the square of that of non-coherent accumulation.
\end{enumerate}

The structure of this paper is organized as follows. Section~\ref{sec:signal_model} presents the signal model for one-bit blind spectrum sensing. In Section~\ref{sec:detector}, a detector based on Rao's test is derived. The null and non-null distributions of the proposed detector are analyzed in Section~\ref{sec:performance}. Section~\ref{sec:comparison} examines the performance degradation when using one-bit ADCs in comparison to $\infty$-bit ADCs. Simulation results are provided in Section~\ref{sec:simulations} to validate the theoretical calculations. Finally, Section~\ref{sec:conclusions} offers a summary of the main conclusions.

\subsection*{Notation}

Throughout this paper, we use boldface uppercase letters for matrices, boldface lowercase letters for column vectors, and light face lowercase letters for scalar quantities. The notation $\A\in\mathbb{R}^{p\times q} \ (\mathbb{C}^{p\times q})$ indicates that $\A$ is a $p\times q$ real (complex) matrix. The operator $||\cdot||$ represents the Frobenius norm when its argument is a matrix, and the $\ell_2$ norm when its argument is a vector. The trace of $\A$ is $\tr(\A)$. The superscripts  $(\cdot)^{-1}$, $(\cdot)^T$, and $(\cdot)^H$ represent matrix inverse, transpose, and Hermitian transpose operations. The operator $\mathbb{E}[a]$ denotes the expected value  and $\sim$ means ``distributed as.'' The central and non-central Chi-squared distributions are denoted by $\chi^2_k$ and $\chi^2_k(\delta^2)$, respectively, where $k$ is the number of degrees-of-freedom (DOFs) and $\delta^2$ is the non-centrality parameter. Finally, the operators $\operatorname{Re}(\cdot)$ and $\operatorname{Im}(\cdot)$ extract the real and imaginary parts of their arguments, $\imath$ is the imaginary unit, and $\mathrm{sign}(\cdot)$ takes the sign of its argument.

\section{Signal Model}
\label{sec:signal_model}

Consider a multiple-input multiple-output CR network where there are $p$ single-antenna primary users (PUs) and $m$ receiving antennas in the secondary user (SU). The input of the one-bit ADCs, $\mathbf{x}(t), t=1,\cdots,n$, under $\mathcal{H}_0$ (signal absence) and $\mathcal{H}_1$ (signal presence) are given by
\begin{align}\label{INF_model}
\begin{array}{l}
\mathcal{H}_{0}:\mathbf{x}(t)=\mathbf{w}(t), \\
\mathcal{H}_{1}: \mathbf{x}(t)=\mathbf{H}\mathbf{s}(t)+\mathbf{w}(t),
\end{array}
\end{align}
where $\mathbf{H} \in \mathbb{C}^{m \times p}$ represents the unknown and deterministic channel coefficient during the sensing period. The signal vector $\mathbf{s}(t)=\left[s_1(t),\cdots,s_p(t)\right]^T$ and the noise vector $\mathbf{w}(t)=\left[w_1(t),\cdots,w_m(t)\right]^T$ follow i.i.d. zero mean circular symmetric complex Gaussian (ZMCSCG) distributions with unknown covariance matrices
$\mathbf{R}_{\mathbf{s}}$ and $\mathbf{R}_{\mathbf{w}}=\diag(\sigma_{w_1},\cdots,\sigma_{w_m}),$ respectively. Moreover, the noise is assumed to be independent of the signal. Clearly, $\mathbf{x}(t)$ follows the ZMCSCG distribution, which is determined by the population covariance matrix (PCM), defined as $\mathbf{R}_{\mathbf{x}}=\mathbb{E}\left[\mathbf{x}(t)\mathbf{x}^H(t)\right]$. Under both hypotheses, the PCM is
\begin{align}
\begin{array}{l}
\mathcal{H}_{0}: \mathbf{R}_{\mathbf{x}}=\mathbf{R}_{\mathbf{w}},\\
\mathcal{H}_{1}: \mathbf{R}_{\mathbf{x}}=\mathbf{H}\mathbf{R}_{\mathbf{s}}\mathbf{H}^H+\mathbf{R}_{\mathbf{w}}.
\end{array}
\end{align}

Before proceeding, we define
\begin{align}
    \mathbf{\tilde{x}}(t)=
\begin{bmatrix}
\Re(\mathbf{x}(t))^T & \Im(\mathbf{x}(t))^T
\end{bmatrix}^T.
\end{align}
Under the circularity assumption, the PCM of $\mathbf{\tilde{x}}(t)$ is given by~\cite{book_peter}:
\begin{align}
    \mathbf{R}_{\mathbf{\tilde{x}}}&=\mathbb{E}\left[\mathbf{\tilde{x}}(t)\mathbf{\tilde{x}}^T(t)\right]\nonumber\\
    &=\frac{1}{2}\begin{bmatrix}
    \Re(\mathbf{R}_{\mathbf{x}})& -\Im(\mathbf{R}_{\mathbf{x}}) \\
     \Im(\mathbf{R}_{\mathbf{x}})&  \Re(\mathbf{R}_{\mathbf{x}})
\end{bmatrix}\in \mathbb{S},
\end{align}
where $\mathbb{S}\subset \mathbb{R}^{2m\times2m}$ is a set of positive definite  matrices of the following form:
\begin{align}
    \begin{bmatrix}
    \mathbf{S}_{1}& -\mathbf{S}_{2} \\
     \mathbf{S}_{2}&  \mathbf{S}_{1}
    \end{bmatrix}.
\end{align}
Here, $\mathbf{S}_{1}\in \mathbb{R}^{m\times m} $ is symmetric and $\mathbf{S}_{2}\in \mathbb{R}^{m\times m}$ is skew-symmetric. On the other hand, considering the diagonal structure of $\mathbf{R}_{\mathbf{x}}$ under $\mathcal{H}_0$, the signal detection problem in \eqref{INF_model} can be rewritten as
\begin{align}\label{INF_model_R}
\begin{array}{l}
\mathcal{H}_{0}: \mathbf{R}_{\mathbf{x}}=\diag(\sigma_{w_1},\cdots,\sigma_{w_m}),\mathbf{R}_{\mathbf{\tilde{x}}}\in \mathbb{S},\\
\mathcal{H}_{1}: \mathbf{R}_{\mathbf{x}}\ne \diag(\sigma_{w_1},\cdots,\sigma_{w_m}),\mathbf{R}_{\mathbf{\tilde{x}}}\in \mathbb{S}.
\end{array}
\end{align}

After one-bit quantization, the output $\mathbf{y}(t)$ is given by
\begin{equation}
    \mathbf{y}(t)=\mathcal{Q}\left(\mathbf{x}(t)\right)=\text{sign}(\Re(\mathbf{x}(t)))+\imath \text{sign}(\Im(\mathbf{x}(t))),
\end{equation}
where $\mathcal{Q}(\cdot)$ represents the one-bit quantization operator. Under each hypothesis, $\mathbf{y}(t)$ becomes
\begin{align}\label{onebit_model}
\begin{array}{l}
\mathcal{H}_{0}: \mathbf{y}(t)=\mathcal{Q}(\mathbf{w}(t)), \\
\mathcal{H}_{1}: \mathbf{y}(t)=\mathcal{Q}(\mathbf{H}\mathbf{s}(t)+\mathbf{w}(t)).
\end{array}
\end{align}
The PMF of $\mathbf{y}(t)$ is given by the orthant probabilities, which are determined only by the coherence matrix \cite{Coherence_book} of $\mathbf{\tilde{x}}(t)$ \cite{Bar2002TAES}. For circular signals, the coherence matrix of $\mathbf{\tilde{x}}(t)$ can be expressed as
\begin{align}\label{P}
{\mathbf{P} }&=\diag(\mathbf{R_{\tilde{x}}})^{-\frac{1}{2} }\mathbf{R_{\tilde{x}}}\diag(\mathbf{R_{\tilde{x}} })^{-\frac{1}{2} }\nonumber\\
&=\begin{bmatrix}
    \Re(\mathbf{P}_{\mathbf{x}})& -\Im(\mathbf{P}_{\mathbf{x}}) \\
     \Im(\mathbf{P}_{\mathbf{x}})&  \Re(\mathbf{P}_{\mathbf{x}})
\end{bmatrix}\in \mathbb{S}\cap \mathbb{T},
\end{align}
where $\mathbf{P}_{\mathbf{x}}=\diag(\mathbf{R_{\mathbf{x}}})^{-\frac{1}{2} }\mathbf{R_{\mathbf{x}}}\diag(\mathbf{R_{\mathbf{x}} })^{-\frac{1}{2} }$ is the coherence matrix of $\mathbf{x}(t)$, and $\mathbb{T}\subset \mathbb{R}^{2m\times 2m}$ is the set of matrices whose diagonal elements are all one. Noticing that $\diag(\Im(\mathbf{P}_{\mathbf{x}}))=\mathbf{0}$, there are a total of $m^2-m$ unknown parameters in $\mathbf{P}$, which are collected in the vector
$\boldsymbol{\theta}\in \mathbb{R}^{(m^2-m) \times 1}$:
\begin{align}\label{theta}
    \boldsymbol{\theta}=[{\rho}_{1,2},\cdots,{\rho}_{m-1,m},{\rho}_{1,2+m},\cdots,{\rho}_{m-1,2m}]^T,
\end{align}
where $\rho_{ij}$ is the $(i,j)$th element of $\mathbf{P}$. Therefore, the signal detection problem is now
\begin{align}\label{onebit_model_R}
\begin{array}{l}
\mathcal{H}_{0}: \bth=\0, {\mathbf{P} }\in \mathbb{S}\cap \mathbb{T},\\
\mathcal{H}_{1}: \bth\neq\0, {\mathbf{P} }\in \mathbb{S}\cap \mathbb{T}.
\end{array}
\end{align}

\section{Derivation of Rao's Test}
\label{sec:detector}

To simplify the computation of the orthant probabilities, we first arrange the real and imaginary components of the observations as
\begin{equation}
    \mathbf{\tilde{Y}}=[\mathbf{\tilde{y}}(1),\cdots,\mathbf{\tilde{y}}(n)],
\end{equation}
where
\begin{equation}
    \mathbf{\tilde{y}}(t)=
\begin{bmatrix}
\Re(\mathbf{y}^T(t)) & \Im(\mathbf{y}^T(t))
\end{bmatrix}^T.
\end{equation}
It is straightforward to show that there are $2^{2m}$ possible values of $\mathbf{\tilde{y}}$, $\mathbf{\tilde{y}}^{\kappa }$ ($\kappa=0,1,\cdots,2^{2m}-1$), where $\mathbf{\tilde{y}}$ is the sample population of $\mathbf{\tilde{y}}(t)$. Next, we define the sets of  $\mathbf{\tilde{x}}$, which is the sample population of $\mathbf{\tilde{x}}(t)$, corresponding to each $\mathbf{\tilde{y}}^{\kappa }$($\kappa=0,1,\cdots,2^{2m}-1$):
\begin{equation}
    \mathbb{X}_{\kappa}=\{\mathbf{\tilde{x}} \in \mathbb{R}^{2m} \mid \diag(\mathbf{\tilde{y}}^{\kappa})\mathbf{\tilde{x}}> \mathbf{0}\},
\end{equation}
Therefore, the probability that $\mathbf{\tilde{y}}(t)=\mathbf{\tilde{y}}^{\kappa}$ is
\begin{multline}
    \Pr\{\mathbf{\tilde{y}}(t)=\mathbf{\tilde{y}}^{\kappa}\} =\Pr\{\mathbf{\tilde{x}}\in\mathbb{X}_{\kappa}\} = \\
    \int_{\mathbb{X}_{\kappa}} \frac{1}{(2\pi)^m \left|\mathbf{P}\right|^{\frac{1}{2}}}e^{-\frac{1}{2}\mathbf{\tilde{x}}^T \mathbf{P}^{-1} \mathbf{\tilde{x}}} \mathrm{d}\mathbf{\tilde{x}}.
\end{multline}
Defining $\mathbf{\zeta}_{\kappa}=\diag(\mathbf{\tilde{y}}_{\kappa})\mathbf{\tilde{x}}$, we have
\begin{align}
    &\Pr\{\mathbf{\tilde{y}}(t)=\mathbf{\tilde{y}}^{\kappa}\} \nonumber\\
    &= \int_{0}^{\infty}\cdots\int_{0}^{\infty} \frac{1}{(2\pi)^m \left|\mathbf{P}\right|^{\frac{1}{2}}}e^{-\frac{1}{2}\mathbf{\zeta}_{\kappa}^T (\mathbf{S}^{\kappa})^{-1} \mathbf{\zeta}_{\kappa}} \mathrm{d}\mathbf{\zeta}_{1} \cdots \mathrm{d}\mathbf{\zeta}_{2m} \nonumber\\
    &= \int_{0}^{\infty}\cdots\int_{0}^{\infty} \frac{1}{(2\pi)^m \left|\mathbf{P}\right|^{\frac{1}{2}}}e^{-\frac{1}{2}\mathbf{\tilde{x}}^T (\mathbf{S}^{\kappa})^{-1} \mathbf{\tilde{x}}} \mathrm{d} \mathbf{\tilde{x}},
\end{align}
where
\begin{equation}
    \mathbf{S}^{\kappa}=\diag(\mathbf{\tilde{y}^{\kappa}})\mathbf{P}\diag(\mathbf{\tilde{y}}^{\kappa}).
\end{equation}
Since $|\mathbf{S}^{\kappa}|=|\mathbf{P}|$, $\Pr\{\mathbf{\tilde{y}}(t)=\mathbf{\tilde{y}}^{\kappa}\}$ can be rewritten as
\begin{equation}
    \Pr\{\mathbf{\tilde{y}}(t)=\mathbf{\tilde{y}}^{\kappa}\}=\phi[\mathbf{S}^{\kappa}],
\end{equation}
where
\begin{equation}
   \phi[\bSi]=\int_{0}^{\infty}\cdots\int_{0}^{\infty} \frac{1}{(2\pi)^m \left|\bSi\right|^{\frac{1}{2}}}e^{-\frac{1}{2}\mathbf{{x}}^T \bSi^{-1} \mathbf{{x}}} \mathrm{d}\mathbf{{x}}
\end{equation}
is the central orthant probability.

Therefore, the likelihood of $\mathbf{\tilde{Y}}$ is
\begin{equation}\label{PMF_Y_h1}
p(\mathbf{\tilde{Y}};\boldsymbol{\theta})=\prod^{n}_{t=1}p(\mathbf{\tilde{y}}(t);\boldsymbol{\theta})=\prod^{n}_{t=1}\phi[{\mathbf{S}}(t)],
\end{equation}
where $p(\mathbf{\tilde{y}}(t);\boldsymbol{\theta})$ is the PMF of $\mathbf{\tilde{y}}(t)$ and
\begin{equation}
    \mathbf{S}(t)=\diag(\mathbf{\tilde{y}}(t))\mathbf{P}\diag(\mathbf{\tilde{y}}(t)).
\end{equation}
 Hence, the log-likelihood function can be expressed as
\begin{equation}\label{log_likelihood}
  \mathcal{L}(\mathbf{\tilde{Y}};\boldsymbol{\theta}) = \sum^{n}_{t=1}\log\left( \phi[\mathbf{{S}}(t)]\right).
\end{equation}

Once we have derived the log-likelihood, the statistic of Rao's test is computed as
\begin{equation}\label{Rao_definition}
T_{\text{R}}=
\left(\!\left.\frac{\partial \mathcal{L}(\mathbf{\tilde{Y}}; \boldsymbol{\theta})}{\partial \boldsymbol{\theta}}\right|_{\boldsymbol{\theta}=\boldsymbol{\theta}_{0}}
\!\right)^{T}\!\!\mathbf{F}^{-1}\!\!\left(\boldsymbol{\theta}_{0}\right)
\left(\left.\frac{\partial \mathcal{L}(\mathbf{\tilde{Y}}; \boldsymbol{\theta})}{\partial \boldsymbol{\theta}}\right|_{\boldsymbol{\theta}=\boldsymbol{\theta}_{0}}
\!\right),
\end{equation}
where $\boldsymbol{\theta}_{0}=\mathbf{0} \in \mathbb{R}^{(m^2-m) \times 1}$ corresponds to the parameters under $\mathcal{H}_0$ and $\mathbf{F}\left(\boldsymbol{\theta}\right)$ is the Fisher information matrix (FIM), which is defined as
\begin{equation}\label{FIM_definition}
\mathbf{F}(\boldsymbol{\theta})=\mathbb{E}\left[ \frac{\partial\mathcal{L}(\mathbf{\tilde{Y}};\boldsymbol{\theta})}{\partial \boldsymbol{\theta}}\frac{\partial\mathcal{L}(\mathbf{\tilde{Y}};\boldsymbol{\theta})}{\partial \boldsymbol{\theta}^T}   \right].
\end{equation}
The result of \eqref{Rao_definition} is provided by the following theorem.
\begin{theorem}\label{theorem:RaoTest}
The Rao's test corresponding to the hypothesis testing problem \eqref{onebit_model_R} is given by
  \begin{align}\label{TR}
    T_{\text{R}}=\frac{n}{2}\sum_{\substack{i,j=1\\i<j}}^{m}\left | \hat{r}_{ij} \right |^2,
\end{align}
where $ \hat{r}_{ij} $ is the $(i,j)$ element of the sample covariance matrix (SCM):
\begin{equation}
  \label{eq:SCM}
  \hat{\mathbf{R}}=\frac{1}{n}\sum_{t=1}^{n}\mathbf{y}(t)\mathbf{y}^H (t).
\end{equation}
\end{theorem}
\begin{IEEEproof}
  See Appendix \ref{appendix:A}.
\end{IEEEproof}

Hence, the detection algorithm based on Rao's test is
\begin{equation}\label{detector}
  T_{\text{R}} \mathop{\gtrless}\limits_{\mathcal{H}_0}^{\mathcal{H}_1} \gamma_{\text{R}},
\end{equation}
where $\gamma_{\text{R}}$ represents the threshold.

Moreover, recall that the second-order EMR detector is
\begin{equation}\label{T_EMR}
    T_{\text{EMR}}=\frac{\frac{1}{m}\left \| \hat{
    \mathbf{R}}  \right \| ^{2}}{ \left(\frac{1}{m}\tr \left(\hat{
    \mathbf{R}}\right)\right)^2} \mathop{\gtrless}\limits_{\mathcal{H}_0}^{\mathcal{H}_1} \gamma_{\text{EMR}}.
\end{equation}
Utilizing the fact that the diagonal elements of $\hat{\mathbf{R}}$ are $2$, we have
\begin{equation}\label{T_EMR'}
    T_{\text{EMR}}=\frac{1}{mn}T_{\text{R}}+1.
\end{equation}
This implies that the Rao's test is equivalent to the EMR test employing the complex-valued SCM, which contrasts with the result in~\cite{Zhaoyuan2021WCL} that is formulated based on the expanded real-valued SCM.

\section{Distributions of Proposed Test}
\label{sec:performance}

In this section, the asymptotic distributions of $T_{\text{R}}$ under $\mathcal{H}_0$ and $\mathcal{H}_1$ are derived. Since $T_{\text{R}}$ is bounded on $[0, nm(m-1)]$,  we can choose a Beta distribution to approximate its distribution, after a proper normalization. The approximation is conducted by first computing the first- and second-order moments of the detector and then matching them with that of the Beta distribution to determine the parameters.

\subsection{Distribution under $\mathcal{H}_0$}

To project the detector to the interval of $[0,1]$, we define a new statistic
\begin{equation}\label{T_R_beta}
T'_{\text{R}}=\frac{1}{nm(m-1)} T_{\text{R}},
\end{equation}
Under $\mathcal{H}_0$, the first and second-order moments of $T'_{\text{R}}$ are given in the following theorem.
\begin{theorem}\label{theorem:1}
Under $\mathcal{H}_0$, $T'_{\text{R}}$ has mean
\begin{equation}\label{mu0}
    \mu_0 =\frac{1}{n},
\end{equation}
and variance
\begin{equation}\label{sigma0}
    \sigma_0^2 =\frac{2(n-1)}{m(m-1)n^3}.
\end{equation}
\end{theorem}
\begin{IEEEproof}
See Appendix \ref{appendix:B}.
\end{IEEEproof}

The cumulative distribution function (CDF) of Beta distribution is
\begin{equation}\label{T'Rcdf}
   F( x; \alpha, \beta)=\frac{\Gamma(\alpha+\beta)}{\Gamma(\alpha) \Gamma(\beta)}\mathrm{B}(x;\alpha, \beta),
\end{equation}
where the incomplete Beta function is
\begin{equation}
B(x ; \alpha, \beta)=\int_{0}^{x} z^{\alpha-1}(1-z)^{\beta-1} \mathrm{d}z,
\end{equation}
and
$\Gamma(x)=\int_0^{+\infty} t^{x-1} e^{-t} \mathrm{~d} t (x>0)$ is the Gamma function. In addition, the mean and variance of a Beta function can be calculated as
\begin{align}\label{beta_moments}
    \mu&=\frac{\alpha}{\alpha+\beta}, &
    \sigma^2&=\frac{\alpha\beta}{(\alpha+\beta)^2(\alpha+\beta+1)}.
\end{align}

% \begin{align}\label{alpha}
%     \alpha_i=\frac{\mu_i(\mu_i-\mu_i^2-\sigma_i^2)}{\sigma_i^2},
% \end{align}
% \begin{align}\label{beta}
%      \beta_i=\frac{(1-\mu_i)(\mu_i-\mu_i^2-\sigma_i^2)}{\sigma_i^2}.
% \end{align}
Matching \eqref{beta_moments} with \eqref{mu0} and \eqref{sigma0}, we obtain the approximated null distribution of $T'_{\text{R}}$:
\begin{equation}\label{null_distrbution}
   \Pr\{T'_{\text{R}}<\gamma\}\approx\frac{\Gamma(\alpha_0+\beta_0)}{\Gamma(\alpha_0) \Gamma(\beta_0)}{\mathrm{B}(\gamma;\alpha_0, \beta_0)},
\end{equation}
where
\begin{align}
    \alpha_0=&\frac{nm(m-1)-2}{2n},\\
     \beta_0=&\frac{(n-1)[nm(m-1)-2]}{2n}.
\end{align}

% Using theorem 2, we can easily find that the  CDF of $T_{\text{R}}$ under  $\mathcal{H}_0$ is
% \begin{align}\label{TR_HO_BETA}
%     F_{T}(t)&=\Pr\{T_{\text{R}}>t\}\nonumber\\
%     &=\Pr\left\{T'_{\text{R}}>\nu t\right\}\nonumber\\
%     &=F\left(\nu t; \alpha_0, \beta_0\right).
% \end{align}

% Hence, for a given threshold $\gamma_{\text{R}}$, the false alarm probability of the proposed detector is
% \begin{align}
%     P_{fa}&=1-F_{T}(\gamma_{\text{R}})\nonumber\\
%     &=1-F\left(\nu \gamma_{\text{R}}; \alpha_0, \beta_0\right).
% \end{align}

% For a given $P_{fa}$, the properly selected threshold is
% \begin{align}
%     \gamma_{\text{R}}=\frac{1}{\nu}F^{-1}\left(1-P_{fa}; \alpha_0, \beta_0\right),
% \end{align}
% where $F^{-1}\left(t; \alpha, \beta\right)$ is the inverse function of $F\left(t; \alpha, \beta\right)$.

\subsection{Distribution under $\mathcal{H}_1$}

Under $\mathcal{H}_1 $, the mean and variance of $T'_{\text{R}}$ are given by the following theorem.
\begin{theorem}\label{theorem:2}
Under $\mathcal{H}_1$, the mean of $T'_{\text{R}}$ is
\begin{equation}
    \mu_1=\frac{1}{2m(m-1)}\sum_{\substack{i,j=1\\i<j}}^{m}g_{ij},
\end{equation}
and the variance of $T'_{\text{R}}$ is
\begin{equation}
    \sigma^2_1=\frac{1}{4m^2(m-1)^2}\sum_{\substack{i,j,k,l=1\\i<j,k<l}}^{m}\left(f_{ijkl}-g_{ijkl}\right),
\end{equation}
where $g_{ij}$, $f_{ijkl},$ and $g_{ijkl}$ are defined in Appendix \ref{appendix:C}.
\end{theorem}
 \begin{IEEEproof}
See Appendix \ref{appendix:C}.
\end{IEEEproof}
Similar to $\mathcal{H}_0$, the CDF of $T'_{\text{R}}$ under  $\mathcal{H}_1$ can be approximated by a Beta distribution as
\begin{equation}\label{non_null_distrbution}
   \Pr\{T'_{\text{R}}<\gamma\}\approx\frac{\Gamma(\alpha_1+\beta_1)}{\Gamma(\alpha_1) \Gamma(\beta_1)}{\mathrm{B}(\gamma;\alpha_1, \beta_1)},
\end{equation}
where
\begin{align}
    \alpha_1&=\frac{\mu_1(\mu_1-\mu_1^2-\sigma_1^2)}{\sigma_1^2},\\
     \beta_1&=\frac{(1-\mu_1)(\mu_1-\mu_1^2-\sigma_1^2)}{\sigma_1^2}.
\end{align}
%  Similar to the case in $\mathcal{H}_0$, we can easily obtain the PDF of $T_{\text{R}}$ by the method of variation replacement which is given by
%  \begin{align}\label{TR_H1_BETA}
%     G_{T}(t)&=\Pr \{T_{\text{R}}>t\}\nonumber\\
%     &=\Pr \left\{T'_{\text{R}}>\nu t\right\}\nonumber\\
%     &=F\left(\nu t; \alpha_1, \beta_1\right).
% \end{align}

% Hence, for a given threshold $\gamma_{\text{R}}$, the detection probability of the proposed detector is
% \begin{align}
% P_d&=1-G({\gamma_{\text{R}}})\nonumber\\
% &=1-F\left(\nu \gamma_{\text{R}}; \alpha_1, \beta_1\right).
% \end{align}

\section{Analysis of Performance Degradation}
\label{sec:comparison}

In this section, we investigate the degradation in detection performance when using one-bit ADCs in comparison to $\infty$-bit ADCs. Note that the $\infty$-bit EMR belongs to the category of sphericity tests, which consider both the independence between the random variables and the equality of their variances. However, due to the loss of amplitude information in the one-bit context, it becomes impossible to compare the variances. %As a consequence, the independence test becomes synonymous with the sphericity test. This effect subsequently reduces the degrees of freedom (DoF) of the null distribution by $m-1$.
Thus, we choose to compare our result with the locally most powerful invariant test (LMPIT) for independence in~\cite{ramirez2013TIT}. In fact, when the SNR is low, the diagonal entries of the covariance matrix tend to be close to each other, resulting in the sphericity test delivering performance nearly identical to that of the independence test, as demonstrated by simulations in~\cite{Huang2015TWC}.

\subsection{$\infty$-bit Case}

The detection problem for $\infty$-bit ADCs is \eqref{INF_model_R}. The LMPIT for this problem is~\cite{ramirez2013TIT}
\begin{equation}\label{infty_distrubution}
    T_{\text{L}}=\frac{n}{2} \tr \left(\left(\mathbf{R}_{\mathbf{x}}\diag (\mathbf{R}_{\mathbf{x}})^{-1}-\mathbf{I}_{m}\right)^2\right) \mathop{\gtrless}\limits_{\mathcal{H}_0}^{\mathcal{H}_1} \gamma_{\text{L}},
\end{equation}
and its asymptotic distribution has been analyzed~\cite{xiao2017approximate}:
\begin{equation}
  \label{eq:distributions}
  T_{\text{L}} \sim \begin{cases}
\chi^2_k, & \text{under}~\mathcal{H}_0,\\
\chi^2_k(\delta_{\infty}^2), &\text{under}~\mathcal{H}_1, \\
\end{cases}
\end{equation}
where $k=m^2-m$  and $\delta_{\infty}^2=n \text{tr}[(\mathbf{P}_{\mathbf{x}}-\mathbf{I}_m)^2] = 2 n \|\boldsymbol{\theta}\|^2$. %\davidsays{Can you confirm that $\text{tr}[(\mathbf{P}_{\mathbf{x}}-\mathbf{I}_m)^2] = 2 \|\boldsymbol{\theta}\|^2$ is correct?}
%By combining \eqref{P} and \eqref{theta}, we have
%\begin{align*}
%    \|\boldsymbol{\theta}\|^2&=\frac{1}{2}[\|\Re(\mathbf{P}_{\mathbf{x}})\|^2-\|\diag(\Re(\mathbf{P}_{\mathbf{x}}))\|^2]\\
%    &\quad+\frac{1}{2}[\|\Im(\mathbf{P}_{\mathbf{x}})\|^2-\|\diag(\Im(\mathbf{P}_{\mathbf{x}}))\|^2]\\
%    &=\frac{1}{2}[\|\mathbf{P}_{\mathbf{x}}\|^2- \|\mathbf{I}_m\|^2]\\
%    &=\frac{1}{2}[\|\mathbf{P}_{\mathbf{x}}-\mathbf{I}_m\|^2]\\
%    &=\frac{1}{2}\text{tr}[(\mathbf{P}_{\mathbf{x}}-\mathbf{I}_m)^2]
%\end{align*}

% Noting that
% \begin{align}
%     \mathbf{tr}(\mathbf{P}_{\mathbf{x}}-\mathbf{I}_m)^2 &=\left\|\mathbf{P}_{\mathbf{x}}-\mathbf{I}_m \right\|^2\nonumber\\
%     &=2 \left\| \boldsymbol{\theta} \right\|^2.
% \end{align}

% Thus, $\delta^2$ can rewrite as
% \begin{align}\label{delta}
%     \delta^2=2n \left\| \boldsymbol{\theta} \right\|^2.
% \end{align}

\subsection{One-bit Case}

In Section \ref{sec:performance}, we exploit the Beta distribution to approximate the distribution of $T_{\text{R}}$. However, it is difficult to use it to compare with the $\infty-$bit detectors to analyze the performance degradation. Therefore, we choose to derive a new approximate distribution of $T_{\text{R}}$ in the low-SNR regime in terms of non-central $\chi^2$ distribution. First, we rewrite $T_{\text{R}}$ as
\begin{equation}
T_{\text{R}}=\left\| \tilde{\r}_ {\text{sc}}\right\|^2,
\end{equation}
where $\tilde{\r}_ {\text{sc}}=\sqrt{\frac{n}{2}}\tilde{\r}$. Here, we remind the reader that $\tilde{\mathbf{r}}=\left[\Re(\hat{\mathbf{r}})^T,\Im(\hat{\mathbf{r}})^T\right]^T$, and
\begin{equation}
	\hat{\mathbf{r}}= [\hat{r}_{1,2},\hat{r}_{1,3},\hat{r}_{2,3},\cdots,\hat{r}_{1,m},\cdots,\hat{r}_{m-1,m}]^T.
\end{equation}
The asymptotic distribution of $ \tilde{\r}_ {\text{sc}}$ is presented next.
\begin{theorem}
\label{a_distribution}
In the low-SNR regime where $\bth$ is of order $\mathcal{O}(n^{-\frac{1}{2}})$, the random vector $\tilde{\r}_ {\text{sc}}$ asymptotically follows a multi-dimensional real Gaussian distribution with mean
 \begin{equation}\label{E_r}
    \mathbb{E}[ \tilde{\r}_ {\text{sc}}]=\frac{2\sqrt{2n}}{\pi}\boldsymbol{\theta}+\mathcal{O}(n^{-\frac{1}{2}}),
\end{equation}
 and covariance matrix
 \begin{equation}\label{R_r}
    \mathbf{R}_{ \tilde{\r}_ {\text{sc}}}= \mathbf{I}_{m^2-m}+\mathcal{O}(n^{-\frac{1}{2}}).
\end{equation}
%\davidsays{I am missing a factor $2$ to compensate the $n/2$. I cannot see where I ``lost it''.}
\end{theorem}
\begin{IEEEproof}
See Appendices \ref{appendix:D} and \ref{appendix:E}.
\end{IEEEproof}

 Using the above results, it is easy to conclude that
\begin{equation}
  \label{eq:distributions_one_bit}
  T_{\text{R}} \sim \begin{cases}
\chi^2_k, & \text{under}~\mathcal{H}_0,\\
\chi^2_k(\delta_1^2), &\text{under}~\mathcal{H}_1, \\
\end{cases}
\end{equation}
where
\begin{align}\label{delta1}
    \delta_1^2 = \frac{8n}{\pi ^2}\left\| \boldsymbol{\theta} \right\|^2=\frac{4}{\pi^2}\delta_{\infty}^2.
\end{align}

Therefore, we can deduce that the performance degradation in the low SNR is approximately $10\log_{10}(\sqrt{\delta_{\infty}^2/\delta_1^2}) \approx 2$ dB. Alternatively, this performance loss can be compensated by increasing the sample support by about $\delta_{\infty}^2/\delta_1^2=\pi^2/4 \approx 2.47$ times.

\begin{remark}
It is worth noting that in~\cite{xiao2022tvt}, the $2$ dB loss requires only $\pi/2 \approx 1.57$ times more samples to compensate. When compared with the results in this paper, it becomes evident that the efficiency of non-coherent accumulation is the square root of that of coherent accumulation.
\end{remark}

\section{Numerical Results}
\label{sec:simulations}

In this section, Monte Carlo experiments are conducted.
Firstly, we compare the proposed one-bit Rao's test with the one-bit EMR~\cite{Zhaoyuan2021WCL}. Subsequently, we assess the accuracy of the detector distribution that we have derived under different SNRs. Finally, we verify our theoretical analysis by demonstrating that the performance degradation is as low as $2$ dB.

We conduct $10^5$ Monte Carlo trials for all experiments.
During each trial, we randomly generated the channel coefficient $\mathbf{H}$ through zero-mean circularly symmetric complex Gaussian distribution, normalized its column vectors, and fixed it for each experiment. The SNR is defined as:
\begin{align}
    \text{SNR}=10\log_{10}\left(\frac{\bar{\sigma}_s^2}{\bar{\sigma}_w^2}\right).
\end{align}
where ${\bar{\sigma}_s^2}=\tr(\mathbf{R}_{\mathbf{s}})/p$ and ${\bar{\sigma}^2_w}=\tr(\mathbf{R}_{\mathbf{w}})/m$.

In addition, we evaluate the accuracy of the approximate distribution by utilizing the Cram\'{e}r-von Mises goodness-of-fit criterion, which is defined as
\begin{equation}\label{Error}
  \epsilon = \frac{1}{K}\sum_{i=1}^K\left|F(\xi_i) - \hat{F}(\xi_i)\right|^2,
\end{equation}
where $K$ is the number of thresholds sampled, $\xi_i$ is the $i$ th threshold value, and $F(\xi_i)$ and $ \hat{F}(\xi_i)$ are empirical and  approximate CDFs, respectively.

\subsection{Detection Performance}

Here, we assess the performance of the proposed detection method by comparing it to the one-bit EMR detector~\cite{Zhaoyuan2021WCL} through receiver operating characteristic (ROC) curves. Recall that the expression of the one-bit EMR~\cite{Zhaoyuan2021WCL}:
\begin{align}
    T_{\text{O}}=1+\frac{1}{m}\sum_{\substack{i,j=1\\i<j}}^{2m}\left | \tilde{r}_{ij} \right |^2,
\end{align}
where $\tilde{r}_{ij}$ represents the $(i,j)$ element of the expanded sample covariance matrix
\begin{align}\hat{\mathbf{R}}_{\tilde{\mathbf{y}}}=\frac{1}{n}\sum_{t=1}^{n}\tilde{\mathbf{y}}(t)\tilde{\mathbf{y}}(t)^T.
\end{align}
As illustrated in Fig. 1, the performance of our detector surpasses that of the one-bit EMR detector. The reason for this improvement is that EMR does not take into account the circular property of the received signal. More specifically, $\tilde{r}_{i,i+m}$($i=1,\cdots,m$) is incorporated into the one-bit EMR detection statistic, whereas the population value corresponding to these elements is actually $0$ due to the circularity of the received signal. This introduces excessive DoFs, which consequently leads to performance degradation.
 \begin{figure}[htbp]
  \centering
        \includegraphics[width=0.9\columnwidth]{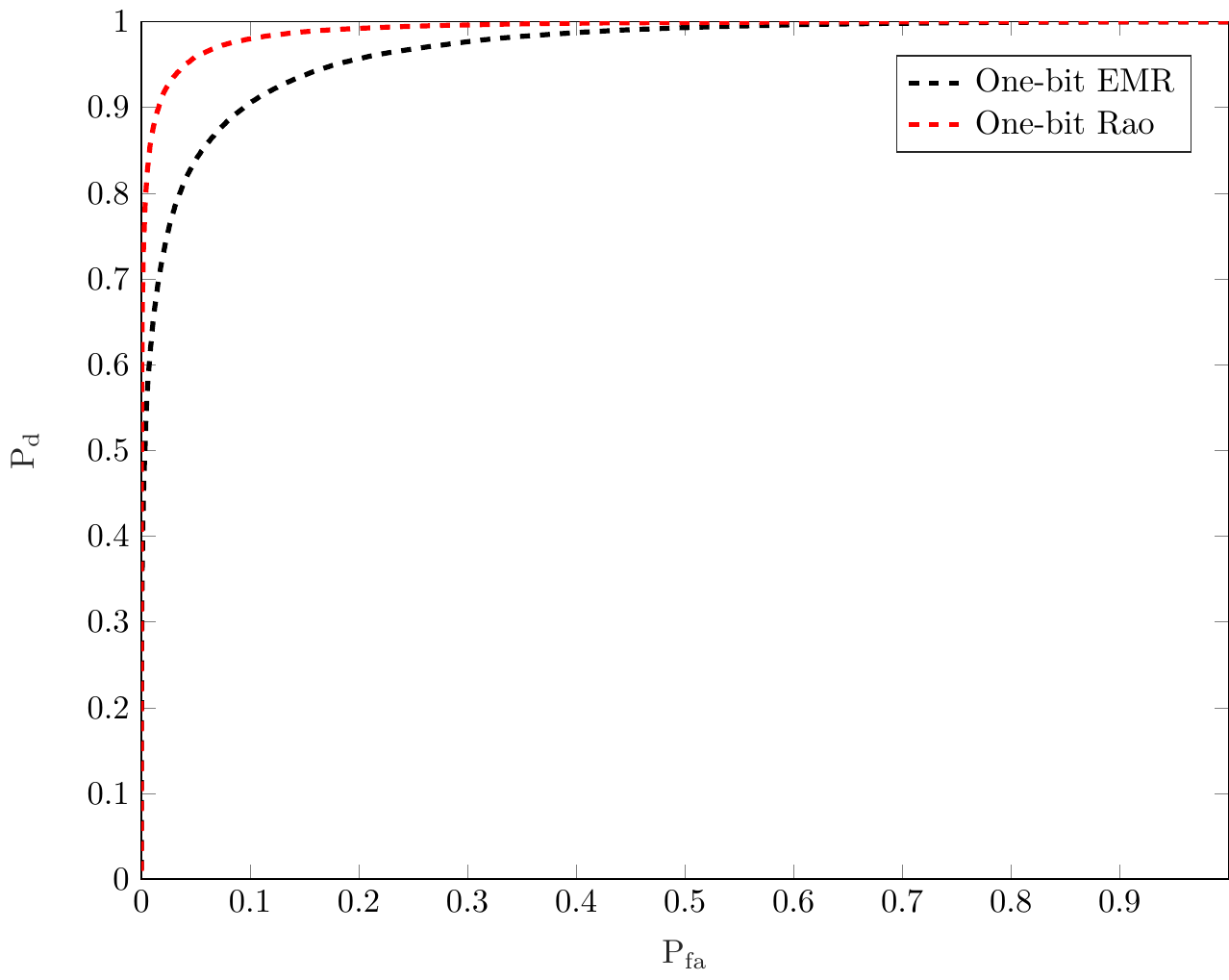}
  \caption{Empirical ROCs for $m=8$, $p=5$, $n=1000$, and SNR$=-7$dB.}\label{fig1}
 \end{figure}

\subsection{Null Distribution}

\begin{figure}[htbp]
  \centering
        \includegraphics[width=0.9\columnwidth]{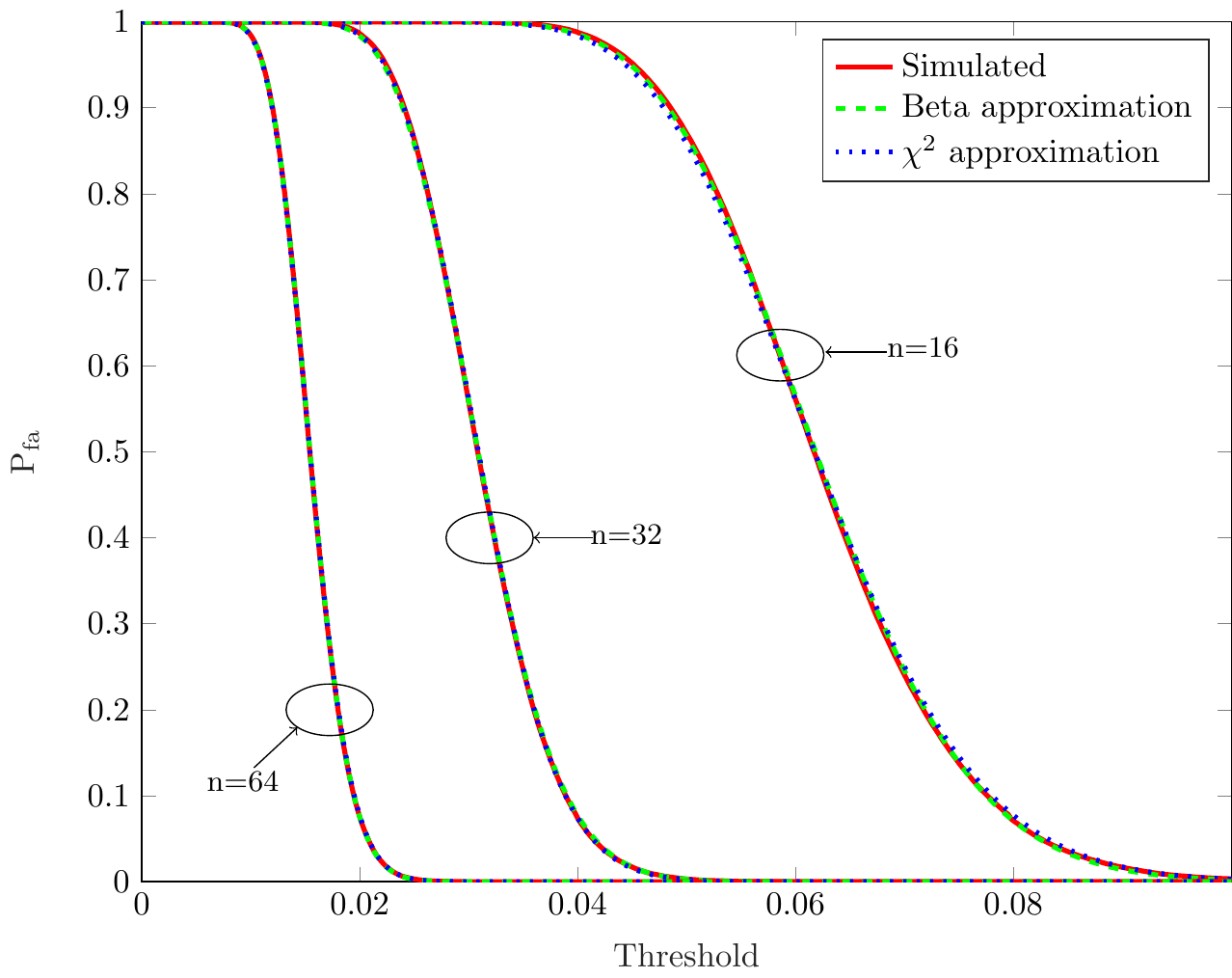}
  \caption{Probability of false alarm versus threshold for $m=8$, $p=5$ and $n=16,32,$ and $64$.}\label{fig2}
 \end{figure}

We first examine the accuracy of the null distribution of the proposed detector. Its approximate distributions include \eqref{null_distrbution} and \eqref{eq:distributions_one_bit}. The simulation results are plotted in Fig. \ref{fig2}. We set $m=8$, $p=5$ and $n=16$, $32$ and $64$. It is worth noting that $T_R'$ is used in in \eqref{T_R_beta}, which belongs to the interval of $[0,1]$. Simultaneously, the result in \eqref{eq:distributions_one_bit} is normalized by setting $\gamma'= \gamma/[mn(m-1)]$. Fig. \ref{fig2}  demonstrates that both the chi-square and beta distributions are capable of fitting the empirical null distribution effectively. This conclusion is substantiated by the approximate error in Table \ref{tab:h0_errors}. Furthermore,  we see that the Beta distribution exhibits a closer approximation to the empirical distribution compared to the $\chi^2$ distribution. An intuitive explanation for this observation is that both the Beta distributions and detection statistics are confined within a specific interval while the $\chi^2$ distribution is not bounded.

\begin{table}[!h]
\begin{center}
\caption{Errors of Null Distribution Approximations.}\label{tab:h0_errors}
\begin{tabular}{ccc}
   \toprule[1.5pt]
   &  $m=8$, $p=5$  \\\midrule[1pt]
  Approximation &~\!\! $n=16$ ~~~~~~~~~~~\!$n=32$ ~~~~~~~~~~\!\! $n=64$\\\midrule[1pt]
   \!\!Eq.~\eqref{null_distrbution} & ~~~$5.44\times10^{-6}$ ~~~ $8.03\times10^{-7}$ ~~~~$1.23\times10^{-7}$ \\
   \!\!Eq.~\eqref{eq:distributions_one_bit} & ~~~$2.09\times10^{-5}$ ~~~ $1.88\times10^{-6}$ ~~~~$3.38\times10^{-7}$\\  \bottomrule[1.5pt]
\end{tabular}
\end{center}
\end{table}

\subsection{Non-null Distribution}

\begin{figure}[!t]
	\begin{minipage}[b]{1\linewidth}
		\centering{\includegraphics[width=0.9\columnwidth]{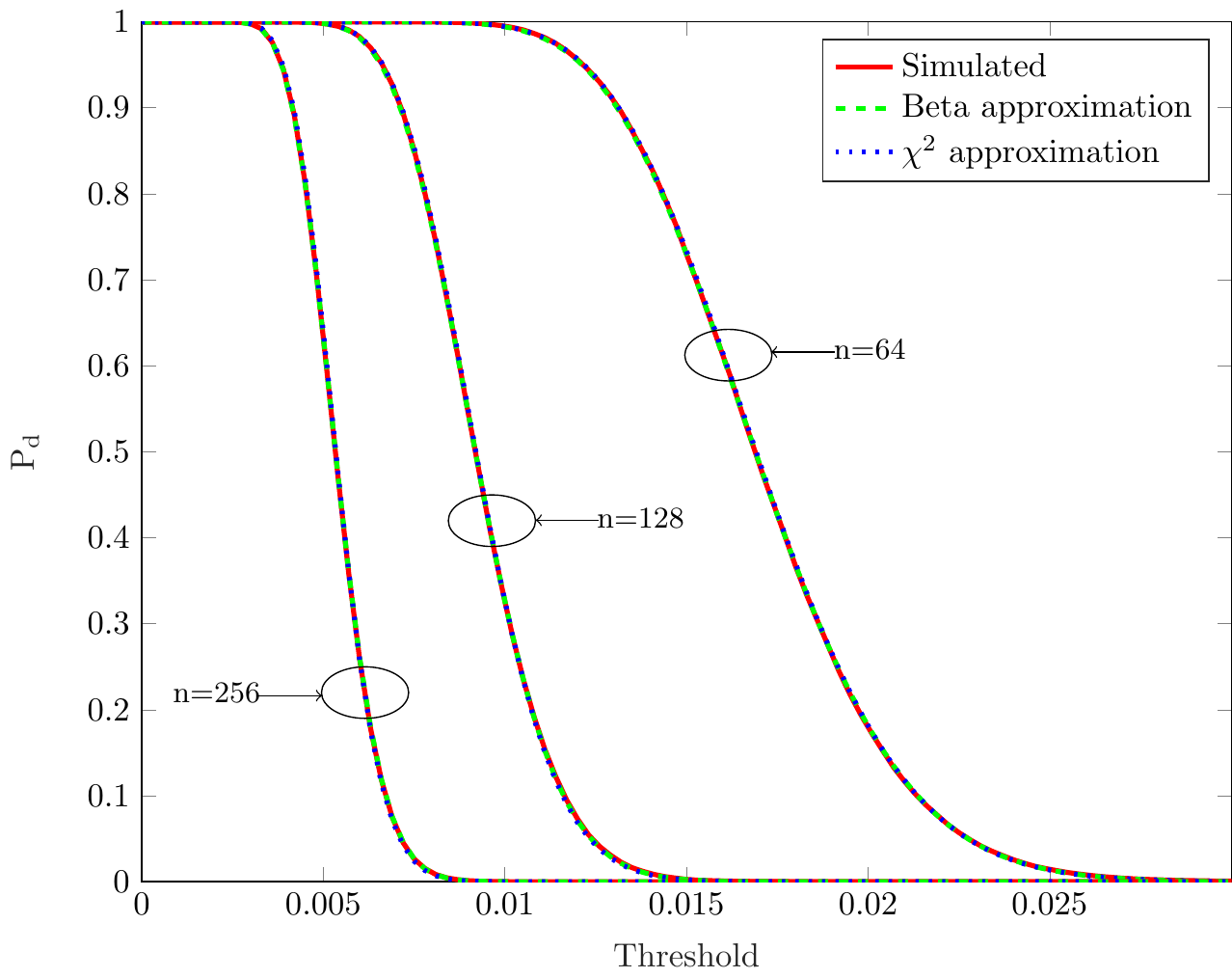}}
		\centerline{\small{(a) SNR=$-5$dB}}
		\medskip
	\end{minipage}
	\hfill
	\begin{minipage}[b]{1\linewidth}
		\centering{\includegraphics[width=0.9\columnwidth]{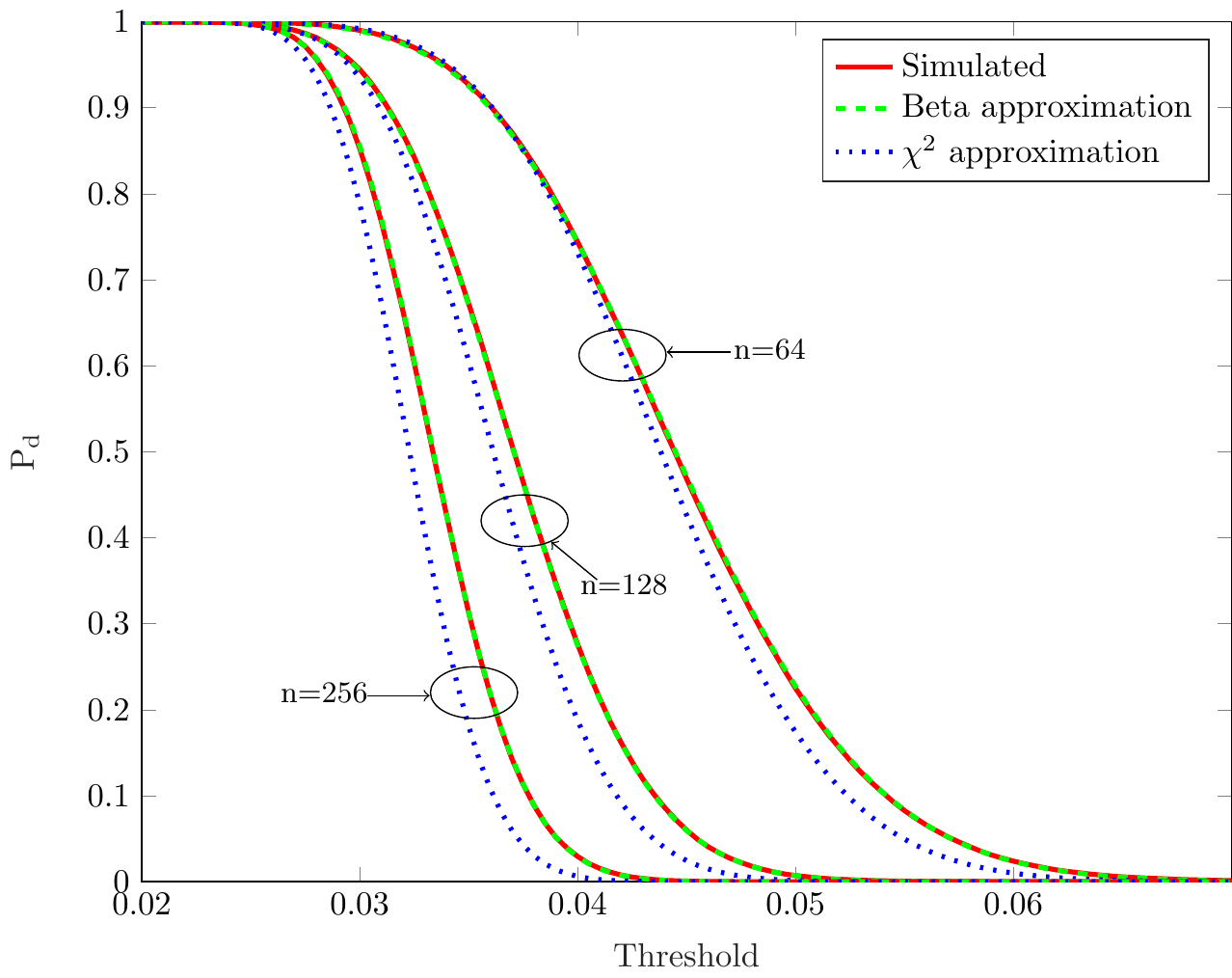}}
		\centerline{\small{(b)  SNR=$5$dB}}
		\medskip
	\end{minipage}
 \vspace{-8mm}
 \caption{Probability of detection versus threshold for $m=8$, $p=5$ and $n=64,128,$ and $256$.}\label{fig3}
\end{figure}

In this section, we investigate the accuracy of the approximate distribution of the proposed detector, which includes \eqref{non_null_distrbution} and \eqref{eq:distributions_one_bit}. The parameters are assigned as $m=8$, $p=5$, $n=64$, $128$, $256$ and $\text{SNR} =-5$dB, $5$dB, with the results displayed in Fig. \ref{fig3}. Fig. \ref{fig3}(a) demonstrates that, in the low SNR regime, both approximations effectively fit the empirical distributions. In contrast, Fig.  \ref{fig3}(b) indicates that, in the high SNR regime, the distribution in \eqref{non_null_distrbution} maintains a good fit to the empirical distributions, whereas the distribution in \eqref{non_null_distrbution} does not exhibit a satisfactory accuracy. The asymptotic errors, as presented in Table II, support these observations.

The reason for these differences in accuracy is that the Beta distribution in~\eqref{non_null_distrbution} is obtained via the method of moments without imposing restrictions on the SNR regime. Conversely, the non-central $\chi^2$ distribution in~\eqref{eq:distributions_one_bit} is derived under the assumption of a low SNR regime, which explains its diminished accuracy in the high SNR context.

%  \begin{figure}[htbp]
%   \centering
%         \includegraphics[width=0.9\columnwidth]{fu15_n64_128_256.pdf}
%   \centerline{\small{(a) SNR=$-15$dB}}
%   \centering
%     \includegraphics[width=0.9\columnwidth]{fu3_n64_128_256.pdf}
%   \centerline{\small{(b) SNR=$-3$dB}}
% %\medskip
%  \caption{Probability of detection versus threshold for $P_{fa}=10^{-2}$, $m=p=16$ and $n=8,32,$ and $128$.}\label{fig3}
%  \end{figure}

% \begin{figure}[htbp]
% \begin{minipage}
%   \centering{\includegraphics[width=0.9\columnwidth]{fu15_n64_128_256.pdf}}
% \centerline{\small{(a) SNR=$-15$d}}
% \medskip
% \end{minipage}
% \hfill
% \begin{minipage}
%   \centering{\includegraphics[width=0.9\columnwidth]{fu3_n64_128_256.pdf}}
% \centerline{\small{(b) SNR=$-3$dB}}
% \medskip
% \end{minipage}
% \caption{Probability of detection versus threshold for $P_{fa}=10^{-2}$, $m=p=16$ and $n=8,32,$ and $128$.}\label{fig3}
% \end{figure}

\begin{table*}[!t]
\begin{center}
\caption{Errors of Non-null Distribution Approximations at Different SNRs.}\label{tab:h1_errors}
\begin{tabular}{ccc}
   \toprule[1.5pt]
   & $\text{SNR}=-5$ dB, $m=8$, $p=5$ & $\text{SNR}=5$ dB, $m=8$, $p=5$ \\\midrule[1pt]
  Approximation &~\!\! $n=64$ ~~~~~~~~~~~\!$n=128$ ~~~~~~~~~~\!\! $n=256$~~~ & $n=64$ ~~~~~~~~~~ $n=128$ ~~~~~~~~ $n=256$ \\\midrule[1pt]
   \!\!Eq.~\eqref{non_null_distrbution} & ~~~$8.62\times10^{-7}$ ~~~ $7.57\times10^{-7}$ ~~~~$1.52\times10^{-7}$~~~~\!&~~$1.65\times10^{-6}$    ~~~~ $5.43\times10^{-7}$ ~~~~$7.99\times10^{-7}$ \\
   \!\!Eq.~\eqref{eq:distributions_one_bit} & ~~~$1.63\times10^{-6}$ ~~~ $2.41\times10^{-6}$ ~~~~$2.54\times10^{-6}$~~~~\!&~~$5.95\times10^{-4}$     ~~~~ $1.40\times10^{-3}$ ~~~~$2.20\times10^{-3}$ \\  \bottomrule[1.5pt]
\end{tabular}
\end{center}
\end{table*}

\subsection{Performance Degradation}

\begin{figure}[htbp]
  \centering
        \includegraphics[width=0.9\columnwidth]{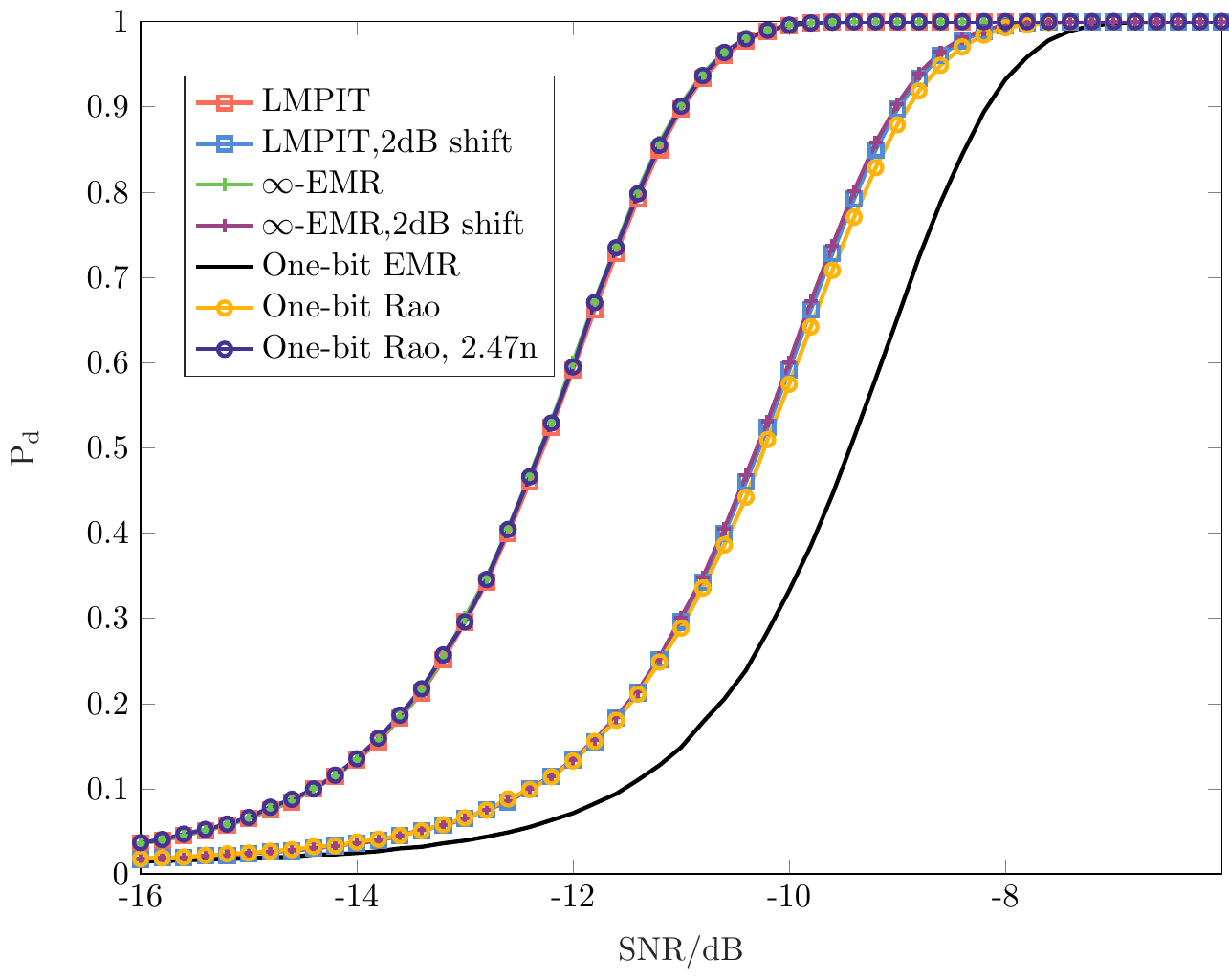}
  \caption{Probability of detection versus SNR for $P_{fa}=10^{-2}$, $m=8$, $p=5$ and $n=2000.$}\label{fig4}
 \end{figure}

In this section, we analyze the performance gap between the proposed one-bit detector and $\infty$-bit detectors. We maintain a fixed false alarm probability of $P_{fa}=0.01$ and examine how the detection probabilities vary with SNR. The parameters are set at $m=8$, $p=5$ and $n=2000$, with the results displayed in Fig.~\ref{fig4}. It can be observed that in the low SNR case and with the same number of samples, the performance degradation of our proposed detector is less than that of the one-bit EMR. Furthermore, the performance degradation of our detector compared to LMPIT and $\infty$-bit EMR is approximately $2$dB, which is consistent with the conclusion we have derived in Section~\ref{sec:comparison}.
On the other hand, Fig.~\ref{fig4} also indicates that the curve of the proposed detector with samples $2.47n$ fits the curves of LMPIT and EMR well, which aligns closely with our theoretical prediction.

% In this section, we study the performance gap between the proposed one-bit detector and $\infty$-bit detectors. We fix the false alarm probability at $P_{fa}=0.01$ and study how the detection probabilities vary with SNR. The parameters are set at $m=p=16$ and $n=512$, the results are shown in Fig.\ref{fig4}. We can see that in the case of low SNR and the same number of samples, The performance degradation of our proposed detectors is less than that of one-bit EMR. Moreover, the performance degradation of our proposed detector compared with LMPIT and $\infty-$bit EMR is about $2$dB, which is consistent with the conclusion we obtained in section \ref{sec:comparison}.

% Then, We study the dependence of the performance compensation on the number of samples. From section \ref{sec:comparison}, the performance degradation of the proposed detector with one-bit ADCs compared with the LMPIT detector with $\infty$-bit ADCs can be compensated by expanding the sampling points of the proposed detector by about $2.47$ times. When we use logarithmic coordinates, it is equivalent to translating LMPIT $\text{log}_{10}(\pi^2/4)\approx0.39$ unit length. Fig. \ref{fig5} shows the the detection probabilities vary with $\log_{10}(n)$ for $P_{fa}=10^{-2}$, $m=p=16$ and $\text{SNR}=-13$dB. We can see that the curve of the proposed detector fits the curves of LMPIT and EMR via 0.39 shift well, which matches well with our theoretical prediction.

\color{black}
\section{Conclusion}
\label{sec:conclusions}

In this paper, a closed-form detector based on Rao's test for blind spectrum sensing utilizing one-bit observations is devised. We derive its null and non-null distributions using the method of moments, allowing us to calculate its false alarm and detection probabilities. Furthermore, the performance degradation of the proposed detector in comparison to the LMPIT detector using $\infty$-bit observations is examined.
Through our analysis, we determine that the performance degradation is reduced from $3$ dB to $2$ dB in the case of low SNR. To compensate for this performance degradation, the sampling number of one-bit observations can be increased by approximately $2.47$ times. Additionally, we find that the effectiveness of non-coherent detection is the square root of coherent detection.

As a future work, this approach can be generalized to one-bit sampling with time-varying thresholds in order to incorporate the diagonal elements of the covariance matrix, which could further enhance the detection performance.

\section*{Acknowledgments}
We acknowledge the assistance of OpenAI's language model, ChatGPT, in proofreading and enhancing the clarity of this manuscript. However, it is important to note that the generation of technical content lies solely with the authors.

\appendices

\section{Proof of Theorem \ref{theorem:RaoTest}}
\label{appendix:A}

To obtain $  T_{\text{R}}$, we first need to compute the partial derivative of the log-likelihood function with respect to the unknown parameters at $\boldsymbol{\theta}=\boldsymbol{\theta}_0=\mathbf{0}$.

Using \eqref{log_likelihood}, it is easy to show that
\begin{align}
    \mathcal{L}(\mathbf{\tilde{Y}} ; \boldsymbol{\theta}=\mathbf{0})&=\sum^{n}_{t=1}\log\left( \phi[\mathbf{I}_{2m}]\right) =\sum^{n}_{t=1}\log\left(\left(\frac{1}{2}\right)^{2m}\right)\nonumber\\
    &=-2mn\log\left(2\right).
\end{align}
Similarly, for $1\le i<j\le m$, we have
\begin{align}
    \mathcal{L}(\mathbf{\tilde{Y}} ; \boldsymbol{\theta}=\boldsymbol{\theta}_{i,j})&=\sum^{n}_{t=1}\log\left(\phi[\mathbf{S}_{ij}(t)]\phi[\mathbf{S}_{i'j'}(t)] \phi[\mathbf{I}_{2m-4}]\right)\nonumber\\
    &=\sum_{t=1}^{n}\log(f_{1}(i,j,t))-(2m-4)n\log\left(2\right),
\end{align}
and
\begin{align}
    \mathcal{L}(\mathbf{\tilde{Y}} ; \boldsymbol{\theta}=\boldsymbol{\theta}_{i',j})&=\sum^{n}_{t=1}\log\left(\phi[\mathbf{S}_{i'j}(t)]\phi[\mathbf{S}_{ij'}(t)] \phi[\mathbf{I}_{2m-4}]\right)\nonumber\\
    &=\sum_{t=1}^{n}\log(f_{2}(i,j,t))-(2m-4)n\log\left(2\right),
\end{align}
where $\{i^{'},j^{'}\}=\{i,j\}+m$, $\boldsymbol{\theta}_{a,b}$ is obtained by zeroing out the elements of $\boldsymbol{\theta}$ except $\rho_{ab}$, $\mathbf{I}_{a}$ is the $a \times a$ identity matrix, and
\begin{equation}
    \mathbf{S}_{ab}(t)=\begin{bmatrix}
  1&\tilde{y}_{a}(t)\tilde{y}_{b}(t)\rho_{ab} \\
  \tilde{y}_{a}(t)\tilde{y}_{b}(t)\rho_{ab}&1
\end{bmatrix}.
\end{equation}
Moreover, $f_{1}\left( i,j,t\right)$ and $f_{2}\left( i,j,t\right)$ are
\begin{align}
    f_{1}\left( i,j,t\right)&=\phi[\mathbf{S}_{ij}(t)]\phi[\mathbf{S}_{i'j'}(t)]\nonumber\\
    &=\left(\frac{1}{4}+\frac{1}{2\pi}\arcsin[\tilde{y}_{i}(t)\tilde{y}_{j}(t)\rho_{ij}]\right)\nonumber\\
    &\quad \times\left(\frac{1}{4}+\frac{1}{2\pi}\arcsin[\tilde{y}_{i'}(t)\tilde{y}_{j'}(t)\rho_{i'j'}]\right)\nonumber\\
    &=\left(\frac{1}{4}+\frac{1}{2\pi}\tilde{y}_{i}(t)\tilde{y}_{j}(t)\arcsin\rho_{ij}\right)\nonumber\\
    &\quad \times\left(\frac{1}{4}+\frac{1}{2\pi}\tilde{y}_{i'}(t)\tilde{y}_{j'}(t)\arcsin\rho_{ij}\right),
\end{align}
and
\begin{align}
    f_{2}\left( i,j,t\right)&=\phi[\mathbf{S}_{i'j}(t)]\phi[\mathbf{S}_{ij'}(t)]\nonumber\\
    &=\left(\frac{1}{4}+\frac{1}{2\pi}\arcsin[\tilde{y}_{i'}(t)\tilde{y}_{j}(t)\rho_{i'j}]\right)\nonumber\\
    &\quad \times\left(\frac{1}{4}+\frac{1}{2\pi}\arcsin[\tilde{y}_{i}(t)\tilde{y}_{j'}(t)\rho_{ij'}]\right)\nonumber\\
    &=\left(\frac{1}{4}+\frac{1}{2\pi}\tilde{y}_{i'}(t)\tilde{y}_{j}(t)\arcsin\rho_{i'j}\right)\nonumber\\
     &\quad \times\left(\frac{1}{4}+\frac{1}{2\pi}\tilde{y}_{i}(t)\tilde{y}_{j'}(t)\arcsin(-\rho_{i'j})\right),
\end{align}
where we have used that each $\phi[\mathbf{S}_{ab}(t)]$ is the integral in the positive quadrant of a zero-mean bi-dimensional Gaussian with covariance matrix $\mathbf{S}_{ab}(t)$~\cite{owen2004orthant}.

Using the definition of partial derivative, it is easy to show that
\begin{align}\label{derivative_1}
    \left.\frac{\partial \mathcal{L}(\mathbf{\tilde{Y}} ; \boldsymbol{\theta})}{\partial \rho_{ij}}\right|_{\boldsymbol{\theta}=\boldsymbol{0}}&= \lim_{\rho_{ij} \to 0} \frac{\mathcal{L}(\mathbf{\tilde{Y}} ; \boldsymbol{\theta}=\boldsymbol{\theta}_{i,j})-\mathcal{L}(\mathbf{\tilde{Y}} ; \boldsymbol{\theta}=\mathbf{0})}{\rho_{ij}}\nonumber\\
    &=\frac{2}{\pi}\sum_{t=1}^{n}\left(\tilde{y}_{i}(t)\tilde{y}_{j}(t)+\tilde{y}_{i'}(t)\tilde{y}_{j'}(t)\right) \nonumber\\
    &=\frac{2n}{\pi}\Re(\hat{r}_{ij}),
\end{align}
and
\begin{align}\label{derivative_2}
    \left.\frac{\partial \mathcal{L}(\mathbf{\tilde{Y}} ; \boldsymbol{\theta})}{\partial \rho_{i'j}}\right|_{\boldsymbol{\theta}=\boldsymbol{0}}&= \lim_{\rho_{i'j} \to 0} \frac{\mathcal{L}(\mathbf{\tilde{Y}} ; \boldsymbol{\theta}=\boldsymbol{\theta}_{i',j})-\mathcal{L}(\mathbf{\tilde{Y}} ; \boldsymbol{\theta}=\mathbf{0})}{\rho_{i'j}}\nonumber\\
    &=\frac{2}{\pi}\sum_{t=1}^{n}\left(\tilde{y}_{i'}(t)\tilde{y}_{j}(t)-\tilde{y}_{i}(t)\tilde{y}_{j'}(t)\right) \nonumber\\
    &=\frac{2n}{\pi}\Im(\hat{r}_{ij}),
\end{align}
where $ \hat{r}_{ij} $ is the $(i,j)$ element of the SCM defined in \eqref{eq:SCM}, and we have used L'H$\hat{\text{o}}$pital's rule. Defining the vector $\hat{\mathbf{r}}\in \mathbb{C}^{\frac{m^2-m}{2}\times 1}$ as
\begin{equation}
    \hat{\mathbf{r}}= [\hat{r}_{1,2},\hat{r}_{1,3},\hat{r}_{2,3},\cdots,\hat{r}_{1,m},\cdots,\hat{r}_{m-1,m}]^T,
\end{equation}
we combine \eqref{derivative_1} and \eqref{derivative_2} as
\begin{equation}\label{derivative}
     \left.\frac{\partial \mathcal{L}(\mathbf{\tilde{Y}} ; \boldsymbol{\theta})}{\partial \boldsymbol{\theta}}\right|_{\boldsymbol{\theta}=\boldsymbol{\theta}_{0}}=\frac{2n}{\pi}\tilde{\mathbf{r}},
\end{equation}
where $\tilde{\mathbf{r}}=\left[\Re(\hat{\mathbf{r}})^T,\Im(\hat{\mathbf{r}})^T\right]^T$.

Plugging \eqref{derivative} into \eqref{FIM_definition}, the FIM can be rewritten as
\begin{equation}\label{reFIM}
    \mathbf{F}(\boldsymbol{\theta}_0)=\frac{4n^2}{\pi^2}\mathbb{E}[\tilde{\mathbf{r}}\tilde{\mathbf{r}}^T].
\end{equation}
Under $\mathcal{H}_0$, the PMF of $\tilde{\mathbf{Y}}$ is
\begin{equation}\label{pdf of h0}
    p(\mathbf{\tilde{Y}} ; \boldsymbol{\theta}=\boldsymbol{\theta}_0)=\left(\frac{1}{2}\right)^{2mn},
\end{equation}
allowing the computation of the expected values in \eqref{reFIM}. The expected value  of the real parts is
\begin{align}
    \mathbb{E}[\Re(\hat{r}_{ij})\Re(\hat{r}_{kl})]&=\mathbb{E}\left[\frac{1}{n}\sum_{t=1}^{n}\left(\tilde{y}_{i}(t)\tilde{y}_{j}(t)+\tilde{y}_{i'}(t)\tilde{y}_{j'}(t)\right)\nonumber \right.\\
    &\left.\quad\times \frac{1}{n}\sum_{t=1}^{n}\left(\tilde{y}_{k}(t)\tilde{y}_{l}(t)+\tilde{y}_{k'}(t)\tilde{y}_{l'}(t)\right)\right]\nonumber\\
    &=\frac{2}{n}\delta_{ik}\delta_{jl},
\end{align}
and the expected value of the imaginary parts is
\begin{align}
	\mathbb{E}[\Im(\hat{r}_{ij})\Im(\hat{r}_{kl})]&=\mathbb{E}\left[\frac{1}{n}\sum_{t=1}^{n}\left(\tilde{y}_{i'}(t)\tilde{y}_{j}(t)-\tilde{y}_{i}(t)\tilde{y}_{j'}(t)\right)\nonumber \right.\\
	&\left.\quad\times \frac{1}{n}\sum_{t=1}^{n}\left(\tilde{y}_{k'}(t)\tilde{y}_{l}(t)-\tilde{y}_{k}(t)\tilde{y}_{l'}(t)\right)\right]\nonumber\\
	&=\frac{2}{n}\delta_{ik}\delta_{jl},
\end{align}
while the expected value of product between real and imaginary parts is
\begin{align}
    \mathbb{E}[\Re(\hat{r}_{ij})\Im(\hat{r}_{kl})]&=\mathbb{E}\left[\frac{1}{n}\sum_{t=1}^{n}\left(\tilde{y}_{i}(t)\tilde{y}_{j}(t)+\tilde{y}_{i'}(t)\tilde{y}_{j'}(t)\right)\nonumber \right.\\
    &\left.\quad\times \frac{1}{n}\sum_{t=1}^{n}\left(\tilde{y}_{k'}(t)\tilde{y}_{l}(t)-\tilde{y}_{k}(t)\tilde{y}_{l'}(t)\right)\right]\nonumber\\
    &=0,
\end{align}
where $1\le i<j\le m$ and $1\le k<l\le m$ and $\delta_{ab}$ is the Kronecker delta function. Thus, we have
\begin{equation}
    \mathbb{E}[\tilde{\mathbf{r}}\tilde{\mathbf{r}}^T]=\frac{2}{n}\mathbf{I}_{m^2-m},
\end{equation}
and \eqref{reFIM} becomes
\begin{equation}\label{FIM0}
    \mathbf{F}(\boldsymbol{\theta}_0)=\frac{8n}{\pi^2}\mathbf{I}_{m^2-m}.
\end{equation}
Finally, by substituting \eqref{derivative} and \eqref{FIM0} into \eqref{Rao_definition}, the proof is completed.

\section{Proof of Theorem \ref{theorem:1} \label{appendix:B}}

Since the observations at different times are independent and $\tilde{y}_{a}(t)$ can only be $+1$ or $-1$, we have
\begin{align}\label{expectation_simplify}
    \mathbb{E}\left[\prod_{t=1}^{n}\prod_{a=1}^{2m}(\tilde{y}_{a}(t))^{\eta_{at}}\right]=\prod_{t=1}^{n}\mathbb{E}\left[\prod_{a=1}^{2m}(\tilde{y}_{a}(t))^{\text{mod}(\eta_{at},2)}\right],
\end{align}
where $\eta_{at}\in \mathbb{N}$ and $\text{mod}(\eta,2)$ represents dividing $\eta$ by $2$ and obtaining the remainder.
Based on the PMF of $\mathbf{\tilde{Y}}$, under $\mathcal{H}_0$ in \eqref{pdf of h0}, we can easily get that the elements of $\mathbf{\tilde{Y}}$ are independent of each other and
\begin{align}
    \Pr\{\tilde{y}_{a}(t)=1\}=\Pr\{\tilde{y}_{a}(t)=-1\}=\frac{1}{2}
\end{align}
Therefore, under $\mathcal{H}_0$,  $\mathbb{E}\left[\prod_{t=1}^{n}\prod_{a=1}^{2m}(\tilde{y}_{a}(t))^{\eta_{at}}\right]$ can be calculated as
\begin{align}\label{expectation_simplify_h0}
    \mathbb{E}\left[\prod_{t=1}^{n}\prod_{a=1}^{2m}(\tilde{y}_{a}(t))^{\eta_{at}}\right]&= \prod_{t=1}^{n}\prod_{a=1}^{2m}\mathbb{E}\left[(\tilde{y}_{a}(t))^{\text{mod}(\eta_{at},2)}\right]\nonumber\\
    &=\left\{
\begin{array}{lcl}
1 ,       &      & \text{all}\quad \eta_{at} \quad \text{are even},\\
 0,    &      & \text{otherwise},
\end{array} \right.
\end{align}
Define $z_{ij}(t)=y_{i}(t)y_{j}^{*}(t)$. Using \eqref{expectation_simplify_h0}, we have
\begin{equation}
     \mathbb{E}[z_{ij}(t_1)z_{ij}^{*}(t_2)]=4\delta_{t_1t_2},
\end{equation}
and
\begin{multline}
    \mathbb{E}[z_{ij}(t_1)z_{ij}^{*}(t_2)z_{kl}(t_3)z_{kl}^{*}(t_4)]\\
    = 16\delta_{t_1t_2}\delta_{t_3t_4}+16\delta_{ik}\delta_{jl}\delta_{t_1t_4}\delta_{t_2t_3}(1-\delta_{t_1t_2}\delta_{t_3t_4}),
\end{multline}
% \mathbb{E}[z_{ij}(t_1)z_{ij}^{*}(t_2)]\mathbb{E}[z_{kl}(t_3)z_{kl}^{*}(t_4)]\nonumber\\
where $1 \le i<j \le m$, $1 \le k<l \le m$, $1 \le t_1,t_2,t_3,t_4 \le n$.
Hence, under $\mathcal{H}_0$, the mean of $T'_{\text{R}}$ is
\begin{align}\label{mu0_h0}
    \mu_0 &=\frac{1}{2m(m-1)}\sum_{\substack{i,j=1\\i<j}}^{m}\mathbb{E}\left[\left|\hat{r}_{ij}\right|^2\right]\nonumber\\
     &=\frac{1}{2m(m-1)}\sum_{\substack{i,j=1\\i<j}}^{m}\mathbb{E}\left[\frac{1}{n^2}\sum_{\substack{t_1,t_2=1}}^{n}z_{ij}(t_1)z_{ij}^{*}(t_2)\right]\nonumber\\
    % &=\frac{1}{2m(m-1)}\sum_{\substack{i,j=1\\i<j}}^{m}\frac{1}{n^2}\sum_{\substack{t_1,t_2=1}}^{n}\mathbb{E}\left[z_{ij}(t_1)z_{ij}^{*}(t_2)\right]\nonumber\\
    &=\frac{1}{n},
\end{align}
and its variance can be written as
\begin{equation}\label{sigma0_h0}
    \sigma_0^2 =\frac{1}{4m^2(m-1)^2}\left[\sum_{\substack{i,j,k,l=1\\i<j,k<l}}^{m}\mathbb{E}\left[\lvert \hat{r}_{ij}\rvert^2 \lvert \hat{r}_{kl}\rvert^2\right] \right]-\mu_0^2.
\end{equation}
To compute \eqref{sigma0_h0}, we need to obtain
\begin{align}\label{r_ijkl_h0}
    &\mathbb{E}\left[\lvert \hat{r}_{ij}\rvert^2 \lvert \hat{r}_{kl}\rvert^2\right]\nonumber\\
    &=\mathbb{E}\left[\frac{1}{n^4}\sum_{\substack{t_1,t_2,t_3,t_4=1}}^{n}z_{ij}(t_1)z_{ij}^{*}(t_2)z_{kl}(t_3)z_{kl}^{*}(t_4)\right]\nonumber\\
    &=\frac{1}{n^4}\sum_{\substack{t_1,t_2,t_3,t_4=1}}^{n}\mathbb{E}\left[z_{ij}(t_1)z_{ij}^{*}(t_2)z_{kl}(t_3)z_{kl}^{*}(t_4)\right]\nonumber\\
    &=\frac{1}{n^4}[16n^2+16(n^2-n)\delta_{ik}\delta_{jl}].
\end{align}
Substituting \eqref{mu0_h0} and  \eqref{r_ijkl_h0} into \eqref{sigma0_h0} yields
\begin{align}
    \sigma_0^2&=\frac{1}{4m^2(m-1)^2n^4}\sum_{\substack{i,j,k,l=1\\i<j,k<l}}^{m}[16n^2+16(n^2-n)\delta_{ik}\delta_{jl}]-\frac{1}{n^2}\nonumber\\
    &=\frac{2(n-1)}{m(m-1)n^3}
\end{align}
This completes the proof of Theorem 2.

%Then, we can calculate $\alpha_0$ and $\beta_0$ by substituting \eqref{mu0} and \eqref{sigma0} into \eqref{alpha} and \eqref{beta}, which completes the proof.

%\section{Proof of Theorem \ref{theorem:GLRT}}
%\label{appendix:C}
%
%The likelihood of $\X$ under $\mathcal{H}_1$ is
%\begin{equation}
%%\mathcal{L}(\y;\bth)
%f(\X;\beta,\sigma_n^2) = \frac{1}{\pi^N\sigma_n^{2N}}\exp\[-\frac{\|\X-\beta\Z\|^2_F}{\sigma_n^2}\],
%\end{equation}
%where we can see now that it depends on, both, $\beta$ and $\sigma_n^2$. Under $\mathcal{H}_0$, it is given by $f(\X;\beta = 0,\sigma_n^2)$, and the GLRT therefore becomes
%\begin{equation}
%T_{\text{GLRT}}=\frac{\max_{\sigma_n^2} f(\X;\beta=0,\sigma_n^2)}{\max_{\beta,\sigma_n^2} f(\X;\beta,\sigma_n^2)}  \mathop{\gtrless}\limits_{\mathcal{H}_1}^{\mathcal{H}_0} \gamma.
%\end{equation}
%The MLE of $\beta$ and $\sigma_n^2$ are easily obtained, yielding
%\begin{align}
%\max_{\sigma_n^2} f(\X;\beta=0,\sigma_n^2) &= \frac{N^N}{\pi^N\tr^N(\X\X^H)} \Nn\\%&\frac{n}{\pi^n\sigma_n^{2n}\tr(\X\X^H)} \\
%\max_{\beta,\sigma_n^2} f(\X;\beta,\sigma_n^2) &= \frac{N^N}{\pi^N\|\X-\hat{\beta}\Z\|^{2N}_\text{F}},
%\end{align}
%where
%\begin{equation}
%\hat{\beta} = \frac{\tr(\X\Z^H)}{\tr(\Z\Z^H)} = \frac{\tr(\X\Z^H)}{N}.
%\end{equation}
%Consequently, the GLRT is
%\begin{equation}
%T_{\text{GLRT}} =\frac{\|\X-\hat{\beta}\Z\|_\text{F}^{2N}}{\tr^N(\X\X^H)}=\[1-\frac{|\tr(\X\Z^H)|^2}{N \tr(\X\X^H)}\]^N.
%\end{equation}

\section{Proof of Theorem \ref{theorem:2} \label{appendix:C}}

Under $\mathcal{H}_1$, by combining the closed-form solution of the second and third-order center orthant probabilities in~\cite{owen2004orthant} and \eqref{PMF_Y_h1}, we have the following expected values:
\begin{equation}\label{h_ij}
    h_{ab}=\mathbb{E}[\tilde{y}_{a}(t)\tilde{y}_{b}(t)]=\frac{2}{\pi}\arcsin{\rho_{ab}},
\end{equation}
and
\begin{align}\label{h_ijkl}
    h_{abcd}&=\mathbb{E}[\tilde{y}_{a}(t)\tilde{y}_{b}(t)\tilde{y}_{c}(t)\tilde{y}_{d}(t)]\nonumber\\
    &=16P_{abcd}-1-(h_{ab}+h_{ac}+h_{ad}+h_{bc}+h_{bd}+h_{cd}),
\end{align}
where $1\le a\ne b\ne c\ne d\le 2m$, and
\begin{equation}\label{pabcd}
    P_{abcd}=\Pr\{\tilde{x}_a(t)>0,\tilde{x}_b(t)>0,\tilde{x}_c(t)>0,\tilde{x}_d(t)>0\}.%\nonumber\\
    %&=\phi[\mathbf{\ddot P}_{abcd}]
\end{equation}
The probability $P_{abcd}$, which is the integral in the positive orthant of a $4$-dimensional Gaussian distribution can be computed using the results in~\cite{orthant1964}.
Since $\rho_{ij}=\rho_{i'j'}$ and $\rho_{i'j}=-\rho_{ij'}$, where $1 \le i <j\le m$, we have
\begin{align}\label{h_ij=h_i'j'}
    h_{i'j'}=h_{ij},h_{ij'}=-h_{i'j}
\end{align}
Using \eqref{expectation_simplify}, \eqref{h_ij}, \eqref{h_ijkl}, and \eqref{h_ij=h_i'j'}, we can obtain the expected value of $z_{ij}(t)$:
\begin{equation}\label{muzij}
    \mathbb{E}\left[z_{ij}(t)\right]=2(h_{ij}+\imath h_{i'j}),
\end{equation}
and those of cross-products:
\begin{align}\label{muzijkl}
   &\mathbb{E}\left[z_{ij}(t)z_{kl}(t)\right]\nonumber\\
   &=\left\{
\begin{array}{lcl}
\!\!\!4h_{ii'jj'},       &    \!\!\!\!\!\! \!\!\!   & i=k,j=l,\\
\!\!\!2[h_{ii'jl'}+h_{ii'j'l}+\imath (h_{ii'jl}-h_{ii'j'l'})],       &   \!\!\!\! \!\!\!\! \!   & i=k,j\ne l,\\
\!\!\!4(h_{kj}+\imath h_{k'j}),       &    \!\!\!\!\! \!\!\!\!   & i=l,\\
\!\!\!4(h_{il}+\imath h_{i'l}),        &  \!\!\!\!\!\!  \!\!\!    & j=k,\\
\!\!\!2[h_{jj'ik'}+h_{jj'i'k}+\imath (h_{jj'i'k'}-h_{jj'ik})],      &  \!\! \!\!\!\!\! \! \!   & j=l,i \ne k,\\
\!\!\! \upsilon_1(i,j,k,l)-\upsilon_2(i,j,k,l)\\
\!\!\! +\imath[ \upsilon_3(i,j,k,l) +\upsilon_4(i,j,k,l)],& \!\!\!\! \!\!\!\!\!   &i \ne j \ne k \ne l,
%\!\!\! \sum\limits_{\xi =1}\limits^{2}q_{\xi }(i,j,k,l)-\imath\sum\limits_{\xi =3}\limits^{4}({-1})^{\xi}q_{\xi }(i,j,k,l),    &   \!\!\!\!\! \! \!\!\!   & i \ne j \ne k \ne l,
\end{array} \right.
\end{align}
and
\begin{align}\label{muzijkl_star}
    &\mathbb{E}\left[z_{ij}(t)z^{*}_{kl}(t)\right]\nonumber\\
    &=\left\{
\begin{array}{lcl}
\!\!\!4,       &    \!\!\!\! \!\!\!\!\!  & i=k,j=l,\\
\!\!\!4(h_{lj}+\imath h_{l'j}),       &   \!\!\!\! \!\!\!\!\!   & i=k,j\ne l,\\
\!\!\!2[h_{ii'jk'}+h_{ii'j'k}+\imath (h_{ii'jk}-h_{ii'j'k'})],       &   \!\!\!\! \!\!\!\!\!   & i=l,\\
\!\!\!2[h_{jj'il'}+h_{jj'i'l}+\imath (h_{jj'i'l'}-h_{jj'il})],        &  \!\!\!\! \!\!\!\! \!   & j=k,\\
\!\!\!4(h_{ik}+\imath h_{i'k}),     &   \!\!\!\! \!\!\!\!\!   & j=l,i \ne k,\\
 \!\!\!
 \upsilon_1(i,j,k,l)+\upsilon_2(i,j,k,l)\\
 \!\!\!+\imath[ \upsilon_3(i,j,k,l) -\upsilon_4(i,j,k,l)],& \!\!\!\! \!\!\!\!\!   &i \ne j \ne k \ne l,
% -\sum\limits_{\xi =1}\limits^{2}({-1})^{\xi}q_{\xi }(i,j,k,l)+\imath\sum\limits_{\xi =3}\limits^{4}q_{\xi }(i,j,k,l)    &  \!\!\!\! \!\!\!\! \!   & i \ne j \ne k \ne l
\end{array} \right.
\end{align}
where $1 \le i<j \le m$, $1 \le k<l \le m$, $1 \le t \le n$,
\begin{subequations}
	\begin{align}
     % q^{(1)}_{ijkl}=h_{ijkl}+h_{ijk'l'}+h_{i'j'kl}+h_{i'j'k'l'}\\
     % q^{(2)}_{ijkl}=h_{i'jk'l}-h_{i'jkl'}-h_{ij'k'l}+h_{ij'kl'}\\
     % q^{(3)}_{ijkl}=h_{ijk'l}-h_{ijkl'}+h_{i'j'k'l}-h_{i'j'kl'}\\
     % q^{(4)}_{ijkl}=h_{i'jkl}+h_{i'jk'l'}-h_{ij'kl}-h_{ij'k'l'}
    \upsilon_1(i,j,k,l)=h_{ijkl}+h_{ijk'l'}+h_{i'j'kl}+h_{i'j'k'l'},\\
    \upsilon_2(i,j,k,l)=h_{i'jk'l}-h_{i'jkl'}-h_{ij'k'l}+h_{ij'kl'},\\
    \upsilon_3(i,j,k,l)=h_{i'jkl}+h_{i'jk'l'}-h_{ij'kl}-h_{ij'k'l'},\\
    \upsilon_4(i,j,k,l)=h_{ijk'l}-h_{ijkl'}+h_{i'j'k'l}-h_{i'j'kl'}.
	\end{align}
\end{subequations}
Therefore, the mean of $T'_{\text{R}}$ under $\mathcal{H}_1$ is
\begin{align}\label{mu1}
    \mu_1&=\frac{1}{2m(m-1)}\sum_{\substack{i,j=1\\i<j}}^{m}\mathbb{E}\left[\left|\hat{r}_{ij}\right|^2\right]\nonumber\\
    &=\frac{1}{2m(m-1)}\sum_{\substack{i,j=1\\i<j}}^{m}\mathbb{E}\left[\frac{1}{n^2}\sum_{\substack{t=1 \\ \phantom{a}}}^{n}z_{ij}(t)z_{ij}^{*}(t)\right.\nonumber\\
    &\left.\quad \quad+\frac{1}{n^2}\sum_{\substack{t_1,t_2=1\\t_1\ne t_2}}^{n}z_{ij}(t_1)z_{ij}^{*}(t_2)\right]\nonumber\\
    &=\frac{1}{2m(m-1)}\sum_{\substack{i,j=1\\i<j}}^{m}g_{ij},
\end{align}
where
\begin{align}
    g_{ij}&=\frac{4}{n^2}\left[n+A_{n,2}\left(h^2_{ij}+h^2_{i'j}\right)\right].
\end{align}
Here, $A_{{n},{m}}=\frac{{n}!}{{m}!}$ defines the number of permutations. In addition, the variance can be computed as
\begin{align}\label{sigma1}
    \sigma_1^2 &=\frac{1}{4m^2(m-1)^2}\left[\sum_{\substack{i,j,k,l=1\\i<j,k<l}}^{m}\mathbb{E}\left[\lvert \hat{r}_{ij}\rvert^2 \lvert \hat{r}_{kl}\rvert^2\right] \right.\nonumber\\
    &\left.\quad-\sum_{\substack{i,j,k,l=1\\i<j,k<l}}^{m}\mathbb{E}\left[\lvert \hat{r}_{ij}\rvert^2 \right]\mathbb{E}\left[\lvert \hat{r}_{kl}\rvert^2\right]\right].
\end{align}
When $\delta_{t_1t_2}+\delta_{t_3t_4} \ge 1$, or $t_1\ne t_2 \ne t_3 \ne t_4$, using \eqref{expectation_simplify}, we have
\begin{align}
    \label{muzijkl_muzij_muzkl}
    &\mathbb{E}[z_{ij}(t_1)z_{ij}^{*}(t_2)z_{kl}(t_3)z_{kl}^{*}(t_4)]\nonumber\\
    &= \mathbb{E}[z_{ij}(t_1)z_{ij}^{*}(t_2)]\mathbb{E}[z_{kl}(t_3)z_{kl}^{*}(t_4)].
\end{align}

Then, the first term in \eqref{sigma1} can be computed as:
\begin{align}
    &\mathbb{E}\left[\lvert \hat{r}_{ij}\rvert^2 \lvert \hat{r}_{kl}\rvert^2\right]\Nn\\
    &=\frac{1}{n^4}\mathbb{E}\left[\sum_{\substack{t_1,t_2,t_3,t_4=1}}^{n}z_{ij}(t_1)z_{ij}^{*}(t_2)z_{kl}(t_3)z_{kl}^{*}(t_4)\right]\Nn\\
    &=\mathbb{E}\left[\lvert \hat{r}_{ij}\rvert^2 \right]\mathbb{E}\left[\lvert \hat{r}_{kl}\rvert^2\right]\Nn\\
    &\phantom{=}+\frac{1}{n^4}\sum_{\substack{t_1,t_2,t_3,t_4=1\\(t_1,t_2,t_3,t_4)\in \mathbb{T}}}^{n}\mathbb{E}\left[z_{ij}(t_1)z_{ij}^{*}(t_2)z_{kl}(t_3)z_{kl}^{*}(t_4)\right]
    \Nn\\
    &\phantom{=}-\frac{1}{n^4}\sum_{\substack{t_1,t_2,t_3,t_4=1\\(t_1,t_2,t_3,t_4)\in \mathbb{T}}}^{n}\mathbb{E}\left[z_{ij}(t_1)z_{ij}^{*}(t_2)\right]\mathbb{E}\left[z_{kl}(t_3)z_{kl}^{*}(t_4)\right],
\end{align}
where
\begin{equation}
   \mathbb{T}=\{(a,b,c,d)|\delta_{ab}+\delta_{cd} =0,\delta_{ac}+\delta_{ad}+\delta_{bc}+\delta_{bd} \ge 1\}.
\end{equation}

Taking into account \eqref{muzij}, \eqref{muzijkl}, and \eqref{muzijkl_star}, it can be shown that
\begin{equation}
    \mathbb{E}\left[\lvert \hat{r}_{ij}\rvert^2 \lvert \hat{r}_{kl}\rvert^2\right]=\mathbb{E}\left[\lvert \hat{r}_{ij}\rvert^2 \right]\mathbb{E}\left[\lvert \hat{r}_{kl}\rvert^2\right]+f_{ijkl}-g_{ijkl},
\end{equation}
where
\begin{equation}
    g_{ijkl}=\frac{32(n-1)(2n-3)}{n^3}(h_{ij}^2+h_{i'j}^2)(h_{kl}^2+h_{k'l}^2).
\end{equation}
and
\begin{align}
   f_{ijkl}=\left\{
\begin{array}{lcl}
\tau_{1}(i,j),       &      & i=k,j=l,\\
\tau_{2}(i,j,l),       &      & i=k,j\ne l,\\
\tau_{2}(i,j,k),       &      & i=l,\\
\tau_{2}(j,i,l),      &      & j=k,\\
\tau_{2}(j,i,k),      &      & j=l,i \ne k,\\
\tau_{3}(j,i,k,l),     &      & i \ne j \ne k \ne l,
\end{array} \right.
\end{align}
with
\begin{multline}
	\tau_1(i,j) =\frac{16}{n^4}A_{n,2}\left[1+h_{ii'jj'}^2 \right]\\
	+\frac{32}{n^4}A_{n,3}\left[(h^2_{ij}+h^2_{i'j})+h_{ii'jj'}(h_{ij}^2-h_{i'j}^2)\right],
\end{multline}
\begin{align}
    \tau_2(i,j,k)&=\frac{4}{n^4}A_{n,2}[4(h^2_{jk}+h^2_{jk'}) +(h_{ii'jk'}+h_{ii'j'k})^2\nonumber\\
    &\phantom{=}+(h_{ii'jk}-h_{ii'j'k'})^2]\nonumber\\
    &\phantom{=}+\frac{16}{n^4}A_{n,3}[(h_{ii'jk}-h_{ii'j'k'})(h_{ik}h_{i'j}+h_{ij}h_{i'k})\nonumber\\
    &\phantom{=}+(h_{ii'jk'}+h_{ii'j'k})(h_{ik}h_{ij}-h_{i'j}h_{i'k})\nonumber\\
    &\phantom{=}+2h_{jk}(h_{ik}h_{ij}+h_{i'j}h_{i'k})\nonumber\\
    &\phantom{=}+2h_{jk'}(h_{ik}h_{i'j}-h_{ij}h_{i'k})],
\end{align}
and
\begin{align}
	\tau_3(i,j,k,l)&=\frac{2}{n^4}A_{n,2}\sum_{t=1}^{4}\upsilon^2_t(i,j,k,l)\\
	&\phantom{=}+\frac{16}{n^4}A_{n,3}\left[\upsilon_1(i,j,k,l)h_{ij}h_{kl}\right.\Nn\\
	&\phantom{=}+\upsilon_2(i,j,k,l)h_{i'j}h_{k'l}\Nn\\
	&\phantom{=}+\upsilon_3(i,j,k,l)h_{kl}h_{i'j}\Nn\\
	&\phantom{=}+\left.\upsilon_4(i,j,k,l)h_{ij}h_{k'l}\right].
\end{align}
 Hence, the variance of $T'_{\text{R}}$ under $\mathcal{H}_1$ is
\begin{equation}
    \sigma_1^2 =\frac{1}{4m^2(m-1)^2}\sum_{\substack{i,j,k,l=1\\i<j,k<l}}^{m}\left(f_{ijkl}-g_{ijkl}\right),
\end{equation}
which completes the proof of Theorem \ref{theorem:2}.

\section{Mean and covariance matrix of $\tilde{\r}_{\text{sc}}$}
\label{appendix:D}

We derive the mean and covariance matrix of $\tilde{\r}_{\text{sc}}$ and leave the proof of Gaussianity in Appendix \ref{appendix:E}.

For convenience, we define a random vector $\r$ with the subscript $\text{sc}$ to mean scaling as follows:
\begin{align}
    \r_{\text{sc}}=\sqrt{\frac{n}{2}}\r
\end{align}

Then, $\tilde{\r}_{\text{sc}}$ can be rewritten as $\tilde{\r}_{\text{sc}}=\left[\Re(\hat{\mathbf{r}}_{\text{sc}})^T,\Im(\hat{\mathbf{r}}_{\text{sc}})^T\right]^T$, where
\begin{equation}
	\hat{\mathbf{r}}_{\text{sc}}\!=\! \left[(\hat{r}_{1,2})_{\text{sc}},(\hat{r}_{1,3})_{\text{sc}},(\hat{r}_{2,3})_{\text{sc}},\!\cdots\!,(\hat{r}_{1,m})_{\text{sc}},\!\cdots\!,(\hat{r}_{m-1,m})_{\text{sc}}\right]^T\!\!.
\end{equation}

Since $\boldsymbol{\theta}$ is assumed of order $\mathcal{O}(n^{-\frac{1}{2}})$, we can apply a Taylor's approximation to $p(\tilde{\mathbf{y}}(t);\bth)$ around
$\boldsymbol{\theta}_0=\0$, allowing us to write
\begin{align}\label{pdf_tylor}
    p(\tilde{\mathbf{y}}(t);\bth)&=p(\tilde{\mathbf{y}}(t);\bth_0)+
    \left. \bth^T\frac{\partial p(\tilde{\mathbf{y}}(t))}{\partial\bth}\right|_{\bth=\bth_0}+\mathcal{O}(n^{-1})\Nn\\
    &=\frac{1}{2^{2m}}+\frac{1}{2^{2m-1}\pi}\sum_{\substack{i,j=1\\i<j}}^{m}\Re(z_{ij}(t))\rho_{ij}\nonumber\\
    &\quad \quad+ \frac{1}{2^{2m-1}\pi}\sum_{\substack{i,j=1\\i<j}}^{m}\Im(z_{ij}(t))\rho_{i'j}+\mathcal{O}(n^{-1}).
\end{align}
where the elements of $\left.\frac{\partial p(\tilde{\mathbf{y}}(t))}{\partial\bth}\right|_{\bth=\bth_0}$ are obtained similarly to \eqref{derivative_1} and \eqref{derivative_2}. Since $\rho_{i'j'}=\rho_{ij}$, $\rho_{ij'}=-\rho_{i'j}$ and $\diag(\Im(\mathbf{P}_{\mathbf{x}}))=\mathbf{0}$, $p(\tilde{\mathbf{y}}(t);\bth)$ can be rewritten as
%\davidsays{Are you sure about the indexes of the summation below?}
%{\color{red} [The following is for explanation only:
%\begin{align*}
%   & \sum_{\substack{i,j=1\\i<j}}^{m}\Re(z_{ij}(t))\rho_{ij}+\sum_{\substack{i,j=1\\i<j}}^{m}\Im(z_{ij}(t))\rho_{i'j}\\
%    &=\sum_{\substack{i,j=1\\i<j}}^{m}[\tilde{y}_{i}(t)\tilde{y}_{j}(t)\rho_{ij}+\tilde{y}_{i'}(t)\tilde{y}_{j'}(t)\rho_{i'j'}]\\
%    &\quad+\sum_{\substack{i,j=1\\i<j}}^{m}[\tilde{y}_{i'}(t)\tilde{y}_{j}(t)\rho_{i'j}+\tilde{y}_{i}(t)\tilde{y}_{j'}(t)\rho_{ij'}]\\
%    &=\sum_{\substack{i,j=1\\i<j}}^{m}[\tilde{y}_{i}(t)\tilde{y}_{j}(t)\rho_{ij}]+\sum_{\substack{i,j=m\\i<j}}^{2m}[\tilde{y}_{i}(t)\tilde{y}_{j}(t)\rho_{ij}]\\
%    &\quad+\sum_{\substack{i=2}}^{m}\sum_{\substack{j
%    =m+1}}^{m+i-1}[\tilde{y}_{i}(t)\tilde{y}_{j}(t)\rho_{ij}]+\sum_{\substack{i=1}}^{m-1}\sum_{\substack{j
%    =m+i}}^{2m}[\tilde{y}_{i}(t)\tilde{y}_{j}(t)\rho_{ij}]\\
%    &=\sum_{\substack{i,j=1\\i<j}}^{2m}\tilde{y}_{i}(t)\tilde{y}_{j}(t)\rho_{ij}-\sum_{\substack{i=1}}^{m}\tilde{y}_{i}(t)\tilde{y}_{i'}(t)\rho_{ii'}\\
%    &=\sum_{\substack{i,j=1\\i<j}}^{2m}\tilde{y}_{i}(t)\tilde{y}_{j}(t)\rho_{ij}
%\end{align*}
%]}
\begin{equation}
    p(\tilde{\mathbf{y}}(t);\bth)=\frac{1}{2^{2m}}+\frac{1}{2^{2m-1}\pi}\sum_{\substack{i,j=1\\i<j}}^{2m}\tilde{y}_{i}(t)\tilde{y}_{j}(t)\rho_{ij}+\mathcal{O}(n^{-1}).
\end{equation}
From this PMF, it is easy to obtain
\begin{align}
    p(\tilde{y}_{a}(t),\tilde{y}_{b}(t);\bth)%&=\sum_{\substack{\tilde{y}_{e}(t)=\pm 1\\1\le e \le m^2-1\\e\ne a,b}}p(\tilde{\mathbf{y}}(t);\bth)\Nn\\
    =\frac{1}{4}+\frac{1}{2\pi}\tilde{y}_{a}(t)\tilde{y}_{b}(t)\rho_{ab}+\mathcal{O}(n^{-1}),
\end{align}
and
\begin{align}
    p(\tilde{y}_{a}(t),&\tilde{y}_{b}(t),\tilde{y}_{c}(t),\tilde{y}_{d}(t);\bth)\Nn\\
    %&=\sum_{\substack{\tilde{y}_{e}(t)=\pm 1\\1\le e \le 2m\\e\ne a,b,c,d}}p(\tilde{\mathbf{y}}(t);\bth)\Nn\\
    &=\frac{1}{16}+\frac{1}{8\pi}\tilde{y}_{a}(t)\tilde{y}_{b}(t)\rho_{ab}
    +\frac{1}{8\pi}\tilde{y}_{a}(t)\tilde{y}_{c}(t)\rho_{ac}\nonumber\\
    &\phantom{=}+\frac{1}{8\pi}\tilde{y}_{a}(t)\tilde{y}_{d}(t)\rho_{ad}+\frac{1}{8\pi}\tilde{y}_{b}(t)\tilde{y}_{c}(t)\rho_{bc}\nonumber\\
    &\phantom{=}+\frac{1}{8\pi}\tilde{y}_{b}(t)\tilde{y}_{d}(t)\rho_{bd}+\frac{1}{8\pi}\tilde{y}_{c}(t)\tilde{y}_{d}(t)\rho_{cd}\Nn\\
    &\phantom{=}+\mathcal{O}(n^{-1}),
\end{align}
where $1\le a \ne b \ne c\ne d\le 2m$.
As a consequence, we have
\begin{align}
    \mathbb{E}[\tilde{y}_{a}(t)\tilde{y}_{b}(t)]  &=\sum_{\substack{\tilde{y}_{e}(t)=\pm 1\\e= a,b}}\tilde{y}_{a}(t)\tilde{y}_{b}(t)p(\tilde{y}_{a}(t),\tilde{y}_{b}(t);\bth)\nonumber\\
    &=\frac{2}{\pi}\rho_{ab}+\mathcal{O}(n^{-1}),
\end{align}
and
\begin{align}
    &\mathbb{E}[\tilde{y}_{a}(t)\tilde{y}_{b}(t)\tilde{y}_{c}(t)\tilde{y}_{d}(t)] \nonumber\\
    &=\!\sum_{\substack{\tilde{y}_{e}(t)=\pm 1\\e= a,b,c,d}}\!\tilde{y}_{a}(t)\tilde{y}_{b}(t)\tilde{y}_{c}(t)\tilde{y}_{d}(t)p(\tilde{y}_{a}(t),\tilde{y}_{b}(t),\tilde{y}_{c}(t),\tilde{y}_{d}(t);\bth)\nonumber\\
    &=\mathcal{O}(n^{-1}).
\end{align}
When some or all of the indexes $\{a,b,c,d\}$ are identical, $\mathbb{E}[\tilde{y}_{a}(t)\tilde{y}_{b}(t)\tilde{y}_{c}(t)\tilde{y}_{d}(t)]$ can be simplified by \eqref{expectation_simplify}.
Thus, we can show
 \begin{align}
     \mathbb{E}[\Re\left({(\hat{r}_{ij})}_{\text{sc}}\right)]&=\sqrt{\frac{n}{2}}\mathbb{E}[\Re(\hat{r}_{ij})]\nonumber\\
     &=\sqrt{\frac{n}{2}}\mathbb{E} \left[\frac{1}{n}\sum_{t=1}^{n}(\tilde{y}_{i}(t)\tilde{y}_{j}(t)+\tilde{y}_{i'}(t)\tilde{y}_{j'}(t))\right]\nonumber\\
     &=\sqrt{\frac{n}{2}}\left[\frac{2}{\pi}\rho_{ij}+\frac{2}{\pi}\rho_{i'j'}+\mathcal{O}(n^{-1})\right]\nonumber\\
     &=\frac{2\sqrt{2n}}{\pi}\rho_{ij}+\mathcal{O}(n^{-\frac{1}{2}}),
 \end{align}
 and
 \begin{align}
     \mathbb{E}[\Im\left({(\hat{r}_{ij})}_{\text{sc}}\right)]&=\sqrt{\frac{n}{2}}\mathbb{E}[\Im(\hat{r}_{ij})]\nonumber\\
     &=\sqrt{\frac{n}{2}}\mathbb{E}\left[\frac{1}{n}\sum_{t=1}^{n}(\tilde{y}_{i'}(t)\tilde{y}_{j}(t)-\tilde{y}_{i}(t)\tilde{y}_{j'}(t))\right]\nonumber\\
     &=\sqrt{\frac{n}{2}}\left[\frac{2}{\pi}\rho_{i'j}-\frac{2}{\pi}\rho_{ij'}+\mathcal{O}(n^{-1})\right]\nonumber\\
     &=\frac{2\sqrt{2n}}{\pi}\rho_{i'j}+\mathcal{O}(n^{-\frac{1}{2}}),
 \end{align}
and the expected value of $\tilde{\r}$ becomes
\begin{equation}
    \mathbb{E}[\tilde{\r}_{\text{sc}}]=\frac{2\sqrt{2n}}{\pi}\boldsymbol{\theta}+\mathcal{O}(n^{-\frac{1}{2}}).
\end{equation}

Since the observations at different times are independent, for $t_1\ne t_2$, we have
\begin{equation}
     \mathbb{E}[\tilde{y}_{a}(t_1)\tilde{y}_{b}(t_1)\tilde{y}_{c}(t_2)\tilde{y}_{d}(t_2)]=\mathbb{E}[\tilde{y}_{a}(t_1)\tilde{y}_{b}(t_1)]\mathbb{E}[\tilde{y}_{c}(t_2)\tilde{y}_{d}(t_2)].
\end{equation}
To proceed, we need to compute
\begin{align}\label{cij}
   &\mathbb{E}[\Re\left((\hat{r}_{ij})_{\text{sc}}\right) \Re\left((\hat{r}_{kl})_{\text{sc}}\right)]-\mathbb{E}[\Re\left((\hat{r}_{ij})_{\text{sc}}\right)]\mathbb{E}[\Re\left((\hat{r}_{kl})_{\text{sc}}\right)]\nonumber\\
   &=\frac{n}{2}\mathbb{E}[\Re(\hat{r}_{ij}) \Re(\hat{r}_{kl})]-\frac{n}{2}\mathbb{E}[\Re(\hat{r}_{ij})]\mathbb{E}[\Re(\hat{r}_{kl})]\nonumber\\
   &=\frac{1}{2n}\mathbb{E}\left[\sum_{t_1,t_2=1}^{n}\Re(z_{ij}(t_1))\Re(z_{kl}(t_2))\right]\nonumber\\
   &\quad \quad-\frac{1}{2n}\mathbb{E}\left[\sum_{t_1=1}^{n}\Re(z_{ij}(t_1))\right]\mathbb{E}\left[\sum_{t_2=1}^{n}\Re(z_{kl}(t_2))\right]\nonumber\\
   &=\frac{1}{2n}\sum_{t=1}^{n}\mathbb{E}\left[\Re(z_{ij}(t))\Re(z_{kl}(t))\right]\nonumber\\
   &\quad \quad-\frac{1}{2n}\sum_{t=1}^{n}\mathbb{E}\left[\Re(z_{ij}(t))\right]\mathbb{E}\left[\Re(z_{kl}(t))\right]\nonumber\\
   &=\left\{
\begin{array}{lcl}
1 - \frac{8}{\pi^2}\rho_{ij}^2 + \mathcal{O}(n^{-1}),       &      & i=k,j=l,\\
\frac{2}{ \pi}\rho_{jl} - \frac{8}{\pi^2}\rho_{ij}\rho_{il}+\mathcal{O}(n^{-1}),     &      & i=k,j\ne l,\\
\frac{2}{ \pi}\rho_{jk} - \frac{8}{\pi^2}\rho_{ij}\rho_{ki}+\mathcal{O}(n^{-1}),       &      & i=l,\\
\frac{2}{ \pi}\rho_{il} - \frac{8}{\pi^2}\rho_{ij}\rho_{jl}+\mathcal{O}(n^{-1}),       &      & j=k,\\
\frac{2}{\pi}\rho_{ik} - \frac{8}{\pi^2}\rho_{ij}\rho_{kj}+\mathcal{O}(n^{-1}),       &      & j=l,i \ne k,\\
- \frac{8}{\pi^2}\rho_{ij}\rho_{kl}+\mathcal{O}(n^{-1}),     &      & i \ne j \ne k \ne l.
\end{array} \right.
\end{align}
Since $\boldsymbol{\theta}$ is of order $\mathcal{O}(n^{-\frac{1}{2}})$, the result of \eqref{cij} can be rewritten as
\begin{align}
     &\mathbb{E}[\Re\left((\hat{r}_{ij})_{\text{sc}}\right) \Re\left((\hat{r}_{kl})_{\text{sc}}\right)]-\mathbb{E}[\Re\left((\hat{r}_{ij})_{\text{sc}}\right)]\mathbb{E}[\Re\left((\hat{r}_{kl})_{\text{sc}}\right)]\nonumber\\
    &=\left\{
\begin{array}{lcl}
1 + \mathcal{O}(n^{-1}),       &      & i=k,j=l,\\
 \mathcal{O}(n^{-\frac{1}{2}}),    &      & \text{otherwise},
\end{array} \right.
\end{align}
where $1 \le i<j\le m$ and  $1 \le k<l\le m$.
Similarly, we have
\begin{align}
      &\mathbb{E}[\Re\left((\hat{r}_{ij})_{\text{sc}}\right) \Im\left((\hat{r}_{kl})_{\text{sc}}\right)]-\mathbb{E}[\Re\left((\hat{r}_{ij})_{\text{sc}}\right)]\mathbb{E}[\Im\left((\hat{r}_{kl})_{\text{sc}}\right)]\nonumber\\
    &=\mathcal{O}(n^{-\frac{1}{2}}),
\end{align}
and
\begin{align}
     &\mathbb{E}[\Im\left((\hat{r}_{ij})_{\text{sc}}\right) \Im\left((\hat{r}_{kl})_{\text{sc}}\right)]-\mathbb{E}[\Im\left((\hat{r}_{ij})_{\text{sc}}\right)]\mathbb{E}[\Im\left((\hat{r}_{kl})_{\text{sc}}\right)]\nonumber\\
    &=\left\{
\begin{array}{lcl}
1 + \mathcal{O}(n^{-1}),       &      & i=k,j=l,\\
 \mathcal{O}(n^{-\frac{1}{2}}),    &      & \text{otherwise},
\end{array} \right.
\end{align}
    Hence, the covariance matrix of $\tilde{\r}_{\text{sc}}$ for low SNR is
\begin{equation}
    \mathbf{R}_{\tilde{\r}_{\text{sc}}}=  \mathbf{I}_{m^2-m}+\mathcal{O}(n^{-\frac{1}{2}}).
\end{equation}

% Using this result, we can easily obtain that the distribution of $T_{\text{R}}$ in the low SNR asymptotically as the non-central chi-square distribution with DOF $m^2-m$ and non-central parameter is
% \begin{align}
%     \delta_1^2 = \left\| \mathbf{\mu_a} \right\|^2=\frac{8n}{\pi ^2}\left\| \boldsymbol{\theta} \right\|^2
% \end{align}
% This completes the proof of Theorem \ref{TR_chi}.

\section{Proof of Gaussianity of $\tilde{\r}_{\text{sc}}$}
\label{appendix:E}

We prove that $\tilde{\r}_{\text{sc}}$ asymptotically follows a Gaussian distribution, which completes the proof of Theorem \ref{a_distribution}. For this proof, we need the following lemma, which is a multivariate version of the central limit theorem~\cite{Bentkus2005}.

\begin{lemma}\label{lemma1}
Let $\mathbf{s}=\sum_{t=1}^{n} \mathbf{b}_{t}$, where  $\mathbf{b}_{1}, \ldots, \mathbf{b}_{n} \in \mathbb{R}^{d}$  are mutually independent random vectors with zero mean. Then, as $ n \rightarrow \infty $, $\mathbf{s}$ is asymptotically Gaussian distributed with zero mean and covariance matrix  $\mathbf{C}$  if
\begin{equation}\label{s_gaussian}
    \lim _{n \rightarrow \infty} \sum_{t=1}^{n} \mathbb{E}\left[\left\|\mathbf{C}^{-1 / 2} \mathbf{b}_{t}\right\|^{3}\right]=0.
\end{equation}
\end{lemma}

To use Lemma \ref{lemma1}, we first define a new set of variables
\begin{equation}
    \tilde{\mathbf{z}}_{t}=\sqrt{\frac{1}{2n}}[\Re(\mathbf{z}_t)^T,\Im(\mathbf{z}_t)^T]^T,
\end{equation}
where $1\le i\le n$, and
\begin{equation}
    \mathbf{z}_t=[z_{1,2}(t),z_{1,3}(t),\cdots,z_{m-1,m}(t)]^T.
\end{equation}
We also define
\begin{equation}
    \mathbf{b}_{t}=\tilde{\mathbf{z}}_{t}-\mathbb{E}[\tilde{\mathbf{z}}_{t}].
\end{equation}
Using \eqref{muzij}, we have
\begin{equation}
    \mathbb{E}[\tilde{\mathbf{z}}_{t}]=\left(\frac{8}{n\pi^2}\right)^{\frac{1}{2}}\arcsin{\boldsymbol{\theta}},
\end{equation}
where $\arcsin$ applying to its argument in an element-wise manner. We also define
\begin{equation}
	\mathbf{s}=\sum_{t=1}^{n} \mathbf{b}_{t},
\end{equation}
as in the previous lemma, which allows us to write $\tilde{\r}_{\text{sc}}$ in Theorem \ref{a_distribution} as
\begin{equation}
    \tilde{\r}_{\text{sc}}=\mathbf{s}+\mathbb{E}[\tilde{\r}_{\text{sc}}],
\end{equation}
where the mean of $\tilde{\r}_{\text{sc}}$ is given by
\begin{align}
    \mathbb{E}[\tilde{\r}_{\text{sc}}]=\sum_{t=1}^{n}\mathbb{E}[\tilde{\mathbf{z}}_{t}]=\left(\frac{8n}{\pi^2}\right)^{\frac{1}{2}}\arcsin{\boldsymbol{\theta}}.
\end{align}
In addition, $\mathbf{C}$ is equal to $\mathbf{R}_{\tilde{\r}_{\text{sc}}}$ which is the  covariance matrix of $\tilde{\r}_{\text{sc}}$.

Since translation does not change the distribution type of the variables, we only need to prove that $\mathbf{s}$ is asymptotically Gaussian distributed  to complete the proof of Theorem \ref{a_distribution}.

Using the Cauchy-Schwarz inequality, we have
\begin{equation}
    \mathbb{E}\left[\left\|\mathbf{C}^{-1 / 2} \mathbf{b}_{t}\right\|^{3}\right] \le \left\|\mathbf{C}^{-1 / 2}\right\|^{3}\mathbb{E}\left[\left\| \mathbf{b}_{t}\right\|^{3}\right],
\end{equation}
and since $\left\|\mathbf{C}^{-1 / 2}\right\|^{3}$ is bounded, a sufficient condition for \eqref{s_gaussian} is
\begin{equation}
\lim _{n \rightarrow \infty} \sum_{t=1}^{n} \mathbb{E}\left[\left\|\mathbf{b}_{t}\right\|^{3}\right]=0.
\end{equation}
Noticing that
\begin{align}\label{maxb}
    \max \left\|\mathbf{b}_{t}\right\|^2 &\le \max 2\left(\left\|\tilde{\mathbf{z}}_{t}\right\|^2+\left\|\mathbb{E}[\tilde{\mathbf{z}}_{t}]\right\|^2\right)\nonumber\\
    &= \max 2 \left(\frac{m(m-1)}{n}+\frac{8}{n\pi^2}\left\|\arcsin \boldsymbol{\theta}\right\|^2\right)\nonumber\\
    &=\frac{2m(m-1)}{n}+\frac{16}{n\pi^2}\max_{\boldsymbol{\theta}}\left\|\arcsin \boldsymbol{\theta}\right\|^2\nonumber\\
    &=\frac{6m(m-1)}{n},
\end{align}
we have
\begin{align}
    \lim _{n \rightarrow \infty} \sum_{t=1}^{n}
    \mathbb{E}\left[\left\|\mathbf{b}_{t}\right\|^{3}\right] &\le \lim _{n \rightarrow \infty} \sum_{t=1}^{n} \max \left[\left\|\mathbf{b}_{t}\right\|^{3}\right]\nonumber\\
    &=\lim _{n \rightarrow \infty} \sum_{t=1}^{n} \left[\left(\max \left\|\mathbf{b}_{t}\right\|^{2}\right)^{\frac{3}{2}}\right]\nonumber\\
    &\le \left[6m(m-1)\right]^{\frac{3}{2}} \lim _{n \rightarrow \infty} n^{-\frac{1}{2}}\nonumber\\
    &=0.
\end{align}
This completes the proof.

% \section{}
% \label{appendix:F}

% where $1\le a,b,c,d \le 2m$, and $P_{abcd}$  is the orthant probabilities of vector $\tilde{\mathbf{x}}_{abcd}(t)=[\tilde{x}_a(t),\tilde{x}_b(t),\tilde{x}_c(t)\tilde{x}_d(t)]^T$, which is defined as
% \begin{align}\label{pabcd}
%     P_{abcd}&=Pr\{\tilde{x}_a(t)>0,\tilde{x}_b(t)>0,\tilde{x}_c(t)>0,\tilde{x}_d(t)>0\}\nonumber\\
%     &=\int_{0}^{\infty}\cdots\int_{0}^{\infty} \varphi (\tilde{\mathbf{x}}_{abcd}(t),\mathbf{P}_{abcd})\mathrm{d}\mathbf{\tilde{x}}_{abcd}(t)
% \end{align}
% where
% \begin{align}
%     \mathbf{P}_{abcd}=\diag(\mathbf{R}_{{\mathbf{\tilde{x}}_{abcd}}})^{-\frac{1}{2} }\mathbf{R}_{{\mathbf{\tilde{x}}_{abcd}}}\diag(\mathbf{R}_{{\mathbf{\tilde{x}}_{abcd}}})^{-\frac{1}{2} },
% \end{align}
% and
% \begin{align}
%    \varphi (\tilde{\mathbf{x}}_{abcd}(t),\mathbf{P}_{abcd})=  \frac{1}{(2\pi)^2\left|\mathbf{P}_{abcd}\right|^{-\frac{1}{2}}}e^{-\frac{1}{2}\mathbf{\tilde{x}}_{abcd}(t)^T \mathbf{P}_{abcd}^{-1} \mathbf{\tilde{x}}_{abcd}(t)}.
% \end{align}
% Here, $\mathbf{R}_{{\mathbf{\tilde{x}}_{abcd}}}=\mathbb{E}[\mathbf{\tilde{x}}_{abcd}(t)\mathbf{\tilde{x}}_{abcd}(t)^H]$ is the PCM of $\mathbf{\tilde{x}}_{abcd}(t)$.

% In general, the result of $P_{abcd}$ is not given by a closed expression. Therefore, we use the numerical method in~\cite{orthant1964} to calculate $P_{abcd}$.
\bibliographystyle{IEEEtran}

%\bibliography{xbib}

\end{sloppypar}

\end{document}